\documentstyle[preprint,aps,epsfig]{revtex}
\tighten
\begin{document}
\preprint{IFUM 531/TH   July 1996}
\draft
\date{ July 12, 1996}
\title
 {Perturbative renormalization group, exact results and\\
 high temperature series to order 21 for the $N$-vector spin models\\
 on the square lattice\\} 
\author{P. Butera and M. Comi}
\address
{Istituto Nazionale di Fisica Nucleare\\
Dipartimento di Fisica, Universit\`a di Milano\\ 
Via Celoria 16, 20133 Milano, Italy\\
e-mail: butera@milano.infn.it\\
comi@milano.infn.it}

\maketitle
\begin{abstract}
 High temperature expansions
for the susceptibility and  the second correlation moment of 
the classical $N$-vector model (also known as the 
$O(N)$ symmetric Heisenberg classical spin  model)
on the square lattice are extended   from   order $\beta^{14}$ to
$\beta^{21}$ for arbitrary $N$. For the second field 
derivative of the 
susceptibility the series expansion 
is extended from   order $\beta^{14}$ to
$\beta^{17}$.

 For $ -2 \leq  N < 2$, a numerical analysis  
of the series is performed 
 in order to compare the critical exponents 
$\gamma(N)$, $\nu(N)$ and $\Delta(N)$ 
to exact (though nonrigorous) formulas and 
to compute the "dimensionless four point coupling constant"
 $\hat g_r(N)$. 
 For $N > 2$, we present a study of the analiticity properties
of $\chi$, $\xi$ etc. in the complex $\beta$ plane and describe 
a method  to estimate the parameters 
which characterize their low-temperature 
behaviors. We compare our series estimates to the 
predictions of the perturbative renormalization group theory, 
to exact (but nonrigorous or conjectured) formulas and to 
the results of the $1/N$ expansion, always finding a good agreement.
\end{abstract}

\widetext
\section{Introduction}

We have  extended  the high temperature (HT)
 series expansion of  the  zero field susceptibility 
$\chi(N;\beta)$ and of
 the second correlation moment 
$\mu_2(N;\beta)$ to order $\beta^{21}$ 
and  the expansion of the second 
field derivative of the susceptibility 
 $\chi_{4}(N;\beta)$ to order $\beta^{17}$ 
 for the $N$-vector model\cite{st68}  (also known as the  
$O(N)$ symmetric Heisenberg classical spin model) 
on all bipartite  lattices in $d=2,3,4,5,..$ space 
dimensions\cite{bc95,bc951,bc96x}. 
  The series coefficients have been determined 
by using  the vertex
renormalized linked cluster expansion  (LCE) 
method \cite{w74,mck83,lw} and have been expressed as explicit
functions of the spin dimensionality  $N$. 
This calculation pursues and improves our 
previous work\cite{bcm88,bcm90} to a considerable
 extent: it   summarizes into a 
convenient  format 
a large body of information  for an infinite set of universality 
classes and offers further  insights into the properties of the
$N$-vector model by enabling us to vary with 
continuity the crucial parameter $N$ and 
to study how various physical
 quantities depend on   $N$. 
 
Strictly speaking, the $N-$vector model is 
defined only for positive integer $N$.  Therefore it is possible
 to construct infinitely 
many "analytic interpolations" in the variable  
$N$ of the HT coefficients and, 
as a consequence, of the physical quantities. 
 We have performed  the  "natural"
 analytic interpolation by which  the HT coefficients are 
rational functions of $N$. 
 This is the most interesting 
interpolation because it coincides with 
that used in the $1/N$ expansion as well as in the
usual renormalization group (RG) treatments and moreover 
it is  unique 
in the sense of Carlson theorem\cite{boas}.

 Next  interesting step,  on  which 
we will  report elsewhere \cite{bc96x}, is to 
compile tables of HT coefficients analytically interpolated both  
in $N$ and in  the space dimension $d$. 
 The "natural" analytic interpolation, with respect to $d$, 
of the HT coefficients, 
 which is polynomial in $d$ and equivalent\cite{babe} to the
one of the Fisher-Wilson method\cite{wf}, is 
also unique in the sense  above specified. 
We will thus be able to describe accurately the general $(N,d)$ 
universality class, to achieve  also a "view from HT"  of 
presently inevitable RG approximation schemes such as
the $\epsilon$ expansions (at the upper and at the lower critical 
dimension), the $1/N$ and the $1/d$ expansions
 and possibly to gain a more detailed knowledge of their limitations.
As pointed out in Ref. \cite{babe,griffiths} possible violations
of the convexity of the free energy  or of the Lee-Yang property 
 might occur   for 
noninteger values of $N$ or for noninteger 
values of $d$, respectively,
 but, of course, they do not question the long 
known conceptual and analytical 
advantages of treating $N$ or $d$ as continuous parameters.

 This paper is devoted to the study of the $N$-vector model 
on the square lattice and its main result
 reported in the appendices,  is the tabulation
 of  the HT series expansion coefficients.
We have made an effort to   keep our exposition selfcontained,  
in particular by  reporting  all  known HT expansion coefficients 
of $\chi$, $\mu_2$ and $\chi_{4}$ and not only the newly 
computed ones. It is worth noting that  further sizable 
extensions of these series are not too difficult 
and now are   in progress.
Several interesting, but somewhat intricate, 
computational procedures 
that have made this  laborious calculation 
 (and its forthcoming extensions) possible will be illustrated
 elsewhere\cite{bc96x}. 

 We shall also update, refine and extend our 
previous numerical analysis of
$O(\beta^{14})$ series presented in Ref.\cite{bcm90},  
but our discussion 
will be more sketchy whenever there is some overlap 
with that reference.

The paper is organized as follows.
In section II we introduce our notations and definitions.
In section III and IV we present a numerical 
study of our new HT series which
  are now  long enough for a  reliable  assessment of the
 uncertainties of the analysis initiated in \cite{bcm88,bcm90}.

In section III  we examine the range
$ -2 \leq  N < 2$  in which the $N$-vector model is known
  (or expected) to have an ordinary 
power law critical point at finite temperature 
 and we estimate the critical exponents $\gamma(N)$ of the 
susceptibility,  $\nu(N)$ of the correlation length, as well as 
 the gap exponent $\Delta(N)$ from $\chi_{4}$
in order to  compare them with exact (nonrigorous)
formulas proposed some time ago \cite{cardy,nienh,baxter}.
 We do not reanalyse here the $N=2$ 
case (the Kosterlitz-Thouless model) 
  which has been already
studied in Ref. \cite{bcxy}, but we only 
wish to call the readers attention on
a few tiny numerical errors  which have crept  in the   HT 
 series coefficients at orders 17-20 
reported in \cite{bcxy} and were due to an accidental 
contamination of a numerical file in the final stage 
of that calculation.
 In this paper we have corrected  
 such errors, which of course could be annoying to those
 who wish to extend the computation  and  first have to make 
sure that they are able to reproduce correctly the existing data.
 We have  also checked that, 
 being so small, these errors 
  were of no consequence  at all 
either on  the qualitative and on the 
quantitative results of the 
 analysis in Refs.\cite{bcxy,bcxy1,bcxy2,bcxy21,bcxy22}, 
which therefore does not need 
to be repeated  until significantly longer series and/or better
methods of analysis can really offer new insights, for example 
on the questions raised by Ref.\cite{kenna}.

 In section IV we examine  the set of models with  $N > 2$, 
 which are expected to behave quite 
differently \cite{polyakov}. 
In the last two decades	
 their  features 
have been extensively  explored  by various analytical and 
numerical techniques, with the main motivation that they are 
 lattice regularizations of the 
 field theoretic non-linear O(N)-symmetric $\sigma$-models, which 
share the crucial asymptotic freedom property with  four dimensional gauge 
field theories, but are  much easier to study. 
We  update our previous survey of  
the nearby  singularities of $\chi$, $\mu_2 $
(and $\chi_{4}$) in the complex inverse 
temperature $\beta$ plane \cite{bcm90} and we 
still find no indication of any  
 physical critical point at finite real $\beta$. 
On the contrary, we point out that the low temperature behavior 
appears to join smoothly onto the high temperature behavior 
so that several parameters which 
characterize the low temperature  behavior
  can be  computed in terms of HT series and 
full consistency is obtained with the  predictions 
of the perturbative RG. 

We end the paper by
 comparing our conclusions to some related recent works which, 
either by direct stochastic 
simulations or by analytic approximations 
such as the $1/N$ expansion,
also  test and confirm the predictions of RG.

In the appendices we report the closed form expressions for the
 HT  series coefficients of $\chi$, 
$\mu_2$ and $\chi_{4}$ as functions of the spin dimensionality $N$
and their evaluation for a few specific values of $N$.
Electronic files containing these data may be requested from the authors. 
The present tabulation extends and supersedes the one in Ref.
\cite{bcm90} which, unfortunately, is marred by a few misprints.

\section{Definitions and notations}

The Hamiltonian $H$ of the $N$-vector model is:

\begin{equation}
H \{ v \} = -{\frac {1} {2}}  
\sum_{\langle x,x' \rangle } v(x) \cdot v(x').
\label{hamilt} \end{equation}

where $v(x)$ is a  $N$-component 
classical spin of unit length at the lattice site $x$,   
and the sum extends to all nearest neighbor pairs of  sites. 

The HT expansion coefficients of all 
correlation functions 
$C(N;\beta) = \sum_{r} f_r(N) \beta^{r}$
 are simple rational functions of $N$, 
namely $f_r(N)= P_r(N)/ Q_r(N)$
where $P_r(N)$ and $Q_r(N)$ are 
integer coefficient polynomials in the
variable $N$.
 Therefore the same property is true for
the expansion coefficients of the susceptibility

\begin{equation}
\chi(N,\beta) = \sum_x \langle v(0) \cdot v(x) \rangle_c = 
 1+ \sum_{r=1}^\infty a_r(N) \beta^r ; \label{chi} \end{equation}

of the second correlation moment

\begin{equation}
 \mu_{2}(N,\beta)=\sum_x x^2 \langle v(0) \cdot v(x) \rangle_c = 
 \sum_{r=1}^\infty s_r(N) \beta^r 
\end{equation}

of the second field derivative of the 
susceptibility 
\begin{equation}
 \chi_{4}(N,\beta )=  \sum_{x,y,z}
\langle  v(0) \cdot v(x) v(y) \cdot v(z)\rangle_{c}=
{\frac {1} {N}} \Big( -2 + \sum_{r=1}^\infty d_r(N) \beta^r \Big).
\label{chi4}\end{equation}

It should be noticed that our definitions of $\chi$ and $\mu_{2}$ 
differ  by a factor $\frac {1} {N}$ 
and  our definition of $\chi_{4}$ differs  by a factor
$ \frac {3} {N(N+2)}$ from that of Ref.\cite{lw}
 (apart from misprints, also adopted  in Ref.\cite{bcm90} 
as far as $\chi_{4}$ is concerned).

Like in any calculation of HT series, the correctness of the
numerical results is a decisive issue. Our
 confidence on the validity of this work is based not only
on the numerous direct and indirect internal tests  
  passed by our code, but especially on the fact 
that the space dimension $d$
and the spin dimensionality $N$ enter simply as parameters which 
can be varied to produce, always by the same procedure, 
series 
 in complete  agreement with those already available
(sometimes to a higher order) 
 for specific values of $N$  and $d$ and in the
spherical model limit (namely 
the limit as $N \rightarrow \infty$ at fixed
$\tilde \beta=\beta/N$)\cite{st68}.
More precisely let us observe that
from the tabulation in the Appendix 
it appears that $\tilde a_r(N)\equiv N^r a_r(N) $,
  $\tilde s_r(N) \equiv N^r s_r(N)$, 
and  $ \tilde d_r(N) \equiv N^r d_r(N)$
have a finite limit
as $N \rightarrow \infty$, so that the spherical model 
limit of our series coefficients 
 can be immediately read.
The HT coefficients of the susceptibility 
$\chi^{(s)}(\tilde\beta)$ and 
 the second correlation moment $ \mu_{2}^{(s)}(\tilde\beta)$   of
 the spherical model 
 can then be  used to check in our tables    the 
coefficients of the highest power of $N$ in the numerator 
polynomials  of  $a_r(N)$ and $s_r(N)$ respectively. 
 It should be noticed that 
$ \chi_4$ 
is  $O(1/N)$ for $N \rightarrow \infty$,  
 as it is expected because, in the spherical model 
limit,  
  only the two-spin connected 
correlation functions are nonvanishing.
 However the quantity 
$\hat \chi_4(N,\beta) \equiv N\chi_4(N,\beta)$ has a finite
large $N$ limit  which will be 
denoted by $\chi_4^{(s)}(\tilde\beta)$.
  We remind the interested reader that, in Ref.\cite{bcmo},
 we have tabulated  the HT coefficients   
for the spherical model susceptibility $\chi^{(s)}$ on the square 
lattice through order 63, and that the 
 HT coefficients for  $\mu_{2}^{(s)}$   and  $\chi_{4}^{(s)}$ can be 
obtained from the expansion of $\chi^{(s)}$ by using the formulas 
$\mu_{2}^{(s)}(\tilde\beta)=4\tilde\beta (\chi^{(s)}(\tilde\beta))^2$ and 
$\chi_{4}^{(s)}(\tilde\beta) = -2 (\chi^{(s)}(\tilde\beta))^2
(\chi^{(s)}(\tilde\beta)+\tilde\beta \frac{d\chi^{(s)}}
 {d\tilde\beta})$).
 
Similarly also  the $N=0$ limit, 
which corresponds to the self-avoiding 
walk (SAW) model can be easily obtained from our HT series 
after expressing all quantities in terms of $\tilde\beta$.
Again, it is  the quantity  $\hat \chi_4(N,\tilde\beta) $ that has 
a finite limit for $N=0$.  The HT 
 coefficients are then essentially given by the 
constant terms in  the numerators of  $ a_r(N) $,
  $ s_r(N)$, and  $  d_r(N) $, up to simple factors from the 
denominators.

 More (partial) checks  are trivially obtained 
by setting $N=1,2$ or $3$ in 
our formulas and comparing the 
results to the corresponding available 
 expansions cited below. 

Of course, if a  "complete set" of such checks were 
available,  it could be used to reconstruct the whole computation.

It is interesting to recall that, more than two decades ago, 
HT series valid for all 
$N$ and for a general lattice have 
been  computed up to order $\beta^{8}$
( $\beta^{9}$ for loosely packed lattices)
 \cite{stanrew}.  Later on,  in the case of a square lattice,   
the series were extended
through $\beta^{11}$\cite{bcm88}
and then through $\beta^{14}$\cite{lw,bcm90}.
On the other hand, very long expansions, 
 on the square lattice, have been computed recently,  
 both for the susceptibility and for the second correlation
moment  in the special cases $N=0$\cite{macd,conw} 
(the self-avoiding walk (SAW) model), 
and  $N=2$ \cite{bcxy,bcxy2}( the Kosterlitz-Thouless model),   
by highly efficient algorithms, 
whose performance, however, does not 
excel for space dimensionality 
 larger than two  or which cannot be extended to 
other values of $N$.
 More precisely, for the susceptibility, the published series extend 
through orders  $\beta^{43}$ (recently pushed to $\beta^{51}$
\cite{congu})  for $N=0$
and $\beta^{20}$ for $N=2$, and for the second correlation
moment through $\beta^{27}$ and $\beta^{20}$, respectively.
 The longest published expansions of $\chi_{4}$, valid for any  $N$,  
presently extend only   through $\beta^{14}$\cite{bcm90}.
The $N=1$ case (the 
spin 1/2 Ising model),  which is  much simpler because the model 
is partially solved, should be 
considered separately: in this case the available series
for $\chi$ and $\mu_2$ extending to $\beta^{54}$ 
are tabulated in Ref.\cite{nickel80} while  the series for
 $\chi_{4}$\cite{smck} extends only to $\beta^{17}$ .
When our  work  was being completed, 
another calculation valid for any $N$
was announced \cite{campo} for the nearest
neighbor correlation function, $\chi$ and 
$\mu_2$ (but not for $\chi_4$) 
 by  the technique of group character expansion. 
This procedure  seems to be  efficient  only in
2 space dimensions  and to be presently feasible up to order 21 
on the square lattice,
to order 30 on the exagonal lattice 
and to order 15 on the triangular
 lattice\cite{campo2}.
It is reassuring that our general results, although
 obtained by a completely different procedure,
agree   throughout their common extent also with the 
specific cases $N=2,3,4,8$ tabulated in Ref.\cite{campo2}.

\section{ Analysis of the HT series for $N < 2$}

We will now discuss some of the 
information that can be extracted from the 
series and update the analysis first presented in \cite{bcm90}. 

 Let us recall that 
an exact expression for the critical 
exponent $\nu(N)$ within the range $ -2 \leq N < 2$  
has been conjectured in Ref. \cite{cardy} on the 
basis of an approximate RG analysis.
 Later on   the same expression and an analogous one
for the exponent $\eta(N)$ were derived\cite{nienh}  by observing that 
a special $O(N)$ spin model
(assumed to be a faithful representative of this universality class)
can be mapped into a soluble\cite{baxter} loop gas model.

The conjectured exact exponents are 

\begin{equation}
 \nu(N) = \frac {1} {4-2t} \label{niennu} \end{equation}

and
\begin{equation}
\gamma(N) \equiv (2-\eta(N))\nu(N) =  \frac {3 + t^2} {4t(2 - t)}
\label{niengam} \end{equation}

with $ N = -2 \cos( \frac {2\pi} {t} )$
and $1 \leq t \leq 2 $.

The quantities $\chi$, $\xi$, and $ \chi_{4}$ are then 
expected to display, in the 
whole range $ -2 \leq N < 2$, as $\beta \uparrow \beta_c(N)$,
  the conventional power law critical behaviors 
$ \chi \simeq c_{\chi}(N) (\beta_c(N)-\beta)^{-\gamma(N)}$,\quad 
$ \xi \simeq c_{\xi}(N)(\beta_c(N)-\beta)^{-\nu(N)}$,
and 
$ \chi_{4} \simeq -c_{4}(N) (\beta_c(N)-\beta)^{-\gamma(N) - 2\Delta(N)}$,
with  Wegner "confluent" corrections.
 Here $c_{\chi}(N)$, $c_{\xi}(N)$ and $c_{4}(N)$ are (nonuniversal)
critical amplitudes.

In order to test numerically 
the validity of (\ref{niennu}) and (\ref{niengam}), 
we have  estimated $\gamma(N)$ and $\nu(N)$   by 
forming  first order inhomogeneous differential 
approximants (DA) \cite{gutasy}
of the susceptibility $\chi$ and 
of the "second moment" correlation length squared 
$\xi^2=\mu_2/2d\chi$ respectively. We also 
have computed the gap exponent $\Delta(N)$ from $ \chi_{4}$.
Our numerical procedure consists in averaging over  
all estimates from DA's  in the class
selected by the protocol of analysis of Ref.\cite{gut87}
which use
at least 16 series coefficients in the case 
of $\chi$ and $\xi$, and at least 14 coefficients 
in the case of $\chi_{4}$. 
For each value of $N$, we first estimate $\beta_c(N)$ and $\gamma(N)$
from $\chi$, then we use  $\beta_c(N)$ 
to bias the computation of $\nu(N)$ from 
$\xi^2$ and of $\Delta(N)$ from $\chi_4/\chi$.
As a measure of
the uncertainties we have  taken
three times the rms deviation   
of the approximant estimates.

Alternatively, we have assumed the validity of the hyperscaling
  relation 

\begin{equation} 
2\Delta(N) =2\nu(N) +\gamma(N)
\label{ipers} \end{equation}

for $ -2 \leq N < 2$
and used also the series for $ \chi_{4}$ 
in the computation of $\nu(N)$ by resorting to the so called 
"critical point renormalization"\cite{gutasy,hiley}. 
 In this case we have estimated $\nu(N)$ by 
examining the singularity at $z=1$ of the 
series $\sum_{r} h_r(N) z^r$ 
with coefficients $h_r(N)= d_r(N)/t_r(N)$
 where $\chi^2(N;\beta)= \sum_{r} t_r(N) \beta^r$.  
  Similarly the exponent $\Delta(N)$ 
has been determined from the series
 with coefficients $l_r(N)= d_r(N)/a_r(N)$ 
and the exponent $\gamma(N)$
has been obtained in terms of the series for $\mu_2$ and for $\xi^2$.
This procedure does not require the knowledge of $\beta_c(N)$, but
only seventeen term series are 
available for the computation of $\nu(N)$ and  $\Delta(N)$.

In Fig.1 we have reported our results for
 $\gamma(N)$,  $\nu(N)$ and $2\Delta(N)$ versus $N$
 and compared them to the exact formulas  
(\ref{niennu}), (\ref{niengam}) and to  (\ref{ipers}). 

 In the central region of 
the plot, approximately for  $-1 < N < 1.5$, 
both numerical procedures we have 
followed yield very accurate estimates
 agreeing with the exact formulas within a small fraction of a 
percent. Near both ends of the interval $ -2 < N < 2$ 
the agreement deteriorates because the series have to
crossover to  different singularity
structures in order  to exhibit  
either a gaussian behavior for $N=-2$ or a 
Kosterlitz-Thouless behavior for $N=2$.
However, since the  exponent estimates 
always move in the right direction
as the number of series coefficients is increased, 
we  are confident that, in these 
border regions, we are simply facing   
a numerical approximation problem 
rather than  a breakdown of the exact formulas 
(\ref{niennu}),  (\ref{niengam}) and therefore
 we can conclude that their validity  
as well as the validity of the
 hyperscaling relation (\ref{ipers}) are  
  convincingly supported also by our HT series study.

  In the $N=0$ case our expansion for $\hat \chi_{4}$ 
is the longest presently
available and therefore  it is  
worthwhile to update the verification of
  the hyperscaling relation (\ref{ipers}).
If we bias the first order DA's of 
$\hat \chi_{4}/\chi$ using the value 
$\tilde\beta_c(0)=0.3790525(6)$, obtained in Ref.\cite{conw}
from an $O(\beta^{43})$ series for $\chi$,  we get the estimate 
 $\Delta(0)= 1.422(1) $ which 
is within $0.1\%$ of the predicted value 
$\Delta(0)= 91/64 = 1.421875 $. 
Similarly (and with the same bias), from a study of
$\hat \chi_{4}/\chi^2$, we obtain the  estimate:
$2\Delta(0)- \gamma(0)=1.503(9)$  which  
by (\ref{ipers}) and  (\ref{niennu}) should 
be compared to the exact value $2\nu(0)=1.5$. 
By studying directly $\hat \chi_4$, 
 we obtain the estimate $2\Delta(0) +\gamma(0)=4.175(3) $. 
Adding the last two estimates, we 
can conclude  that $\Delta(0)= 1.419(3) $, which is slightly 
less accurate, but perfectly 
compatible with the previous result.

 We can also estimate with fair accuracy the (nonuniversal) 
critical amplitudes of $\hat \chi_4$
 for $N=0$ and $N=1$ which might be useful 
for reference and comparison with other numerical calculations.
Let us recall that in the Ising model 
case  the critical amplitude of the 
susceptibility $c_{\chi}(1)$ has been 
computed exactly to be $c_{\chi}(1)=0.962581732..$
 and that also the amplitudes of the 
first few subleading confluent corrections to scaling
are known \cite{barou}. Since the first 
 confluent corrections
are found to be analytic,  
and indeed it has been argued\cite{aha} that there 
are no irrelevant-variable corrections 
to scaling in the thermodynamic quantities 
of the two-dimensional Ising model,
we expect that we can    rely quite simply on  near diagonal 
Pad\'e approximants (PA's)
 of $(\beta_c- \beta)^{\gamma(1)}\chi$ to obtain
an accurate estimate of $c_{\chi}(1)$, even   from not 
particularly long series.
Indeed, from our $O(\beta^{21})$ expansion, 
we get $c_{\chi}(1)=0.96261(3) $ using as a bias the 
known values of $\gamma(1)$
 and $\beta_c(1)$.
The critical amplitude of $\tilde \chi_4$ has not 
yet been evaluated exactly, 
but since 
the structure of  the confluent 
corrections  to scaling should be similar to that of $\chi$
also this amplitude should  be accurately  estimated.
 Our biased estimate of this quantity  is  $\hat c_{4}(1)=4.378(2)$ 
which  compares well with the  estimate 
$\hat c_{4}(1)=4.37(1)$ from the fifteen 
term series 
of Ref.\cite{bakergr}.
Also in  the $N=0$ case ( now in terms of the variable 
$\tilde\beta$) the structure 
of the confluent corrections is likely to
 be favorable since both 
the long expansion computed 
for $\chi$ in Ref.\cite{conw} and the 
results of extensive stochastic
simulations \cite{li} are consistent with 
an analytic dominant confluent correction.
In this case we get $c_{\chi}(0)=1.0524(8) $ 
and  $\hat c_{4}(0)=6.62(2)$.

In terms of $\chi$, $\xi$, 
and $\hat \chi_{4}$ we can also compute the 
"dimensionless renormalized 
four point coupling  constant" $\hat g_r(N)$ 
 as the value  of
\begin{equation}
\hat g_r(N,\beta)\equiv - \frac{\hat \chi_{4}(N,\beta)} 
{\xi^2(N,\beta) \chi^2(N,\beta)}
\end{equation}
at the critical point $\beta_c(N)$.
 If we assume  that the  inequality  
$\gamma(N) + 2\nu(N)-2\Delta(N) \geq 0 $,
 (rigorously proved to hold as an equality  for $N=1$), 
is also true for any $ -2 \leq N < 2$ , then 
$\hat g_r(N)$ is a bounded (nonnegative) 
universal amplitude combination
whose vanishing  is a sufficient condition 
for gaussian behavior at criticality, or, in lattice field theory 
language, 
for "triviality" of the continuum limit 
theory defined by the critical $N-$vector model\cite{fernandez}. 
Notice that our normalization of 
$\hat g_r(N)$ is the same as the one adopted
 in Ref. \cite{lw} and differs by 
a factor $\frac {N+8} {8\pi N(N+2)}$ from
the normalization traditionally chosen in the 
 field theoretic renormalization group treatments\cite{zinn}.

For $ 0 \leq N \leq 2$  we have estimated $\hat g_r(N)$ by 
evaluating both near diagonal PA's and first order inhomogeneous 
DA's of the series 
for $1/\hat g_r(N,\beta)$
at the critical values  $\beta_c(N)$.  
The two procedures yield  results which are   
perfectly consistent within their numerical uncertainties.
In Fig. 2
 we have reported  our estimates of $\hat g_r(N)$ for various 
values of $0 \leq N < \infty $
 and compared our results to   other 
 computations in the literature. 

For $N=0$ we estimate $\hat g_r(0)=10.53(2)$, 
which agrees well with a previous
estimate  $\hat g_r(0)=10.51(5)$ 
from the  $O(\beta^{14})$ series \cite{lw,bcm90} studied 
in Ref. \cite{campogr}.

For $N=1$, our estimate $\hat g_r(1)=14.693(4)$ 
coincides with the estimate of 
Ref. \cite{nickelsharpe} obtained using
the same number of coefficients, 
but  a rather different method of analysis,
and is consistent with previous estimates 
from an eleven term 
series\cite{bakergr} giving  $\hat g_r(1)=14.67(5)$
 and from a fourteen term 
series \cite{campogr} giving $\hat g_r(1)=14.63(7)$.
 Less precise, but consistent estimates also
 come  from  field theory   (fixed dimension RG)\cite{guillou}
yielding $\hat g_r(1)=15.5(8)$ and from a recent single 
cluster   MonteCarlo (MC) simulation\cite{kim} 
 yielding $\hat g_r(1)=14(2).$.

For $N=2$,  biasing the approximant   
with  the critical inverse temperature 
$\tilde\beta_c(2)=1.118(4)$, we get the estimate 
$\hat g_r(2)=18.3(2)$. This result is  consistent 
both with the determination
$\hat g_r(2)=18.2(2)$ obtained in 
Ref.\cite{campogr} and with the
 MC   measure 
$\hat g_r(2)=17.7(5)$ obtained in Ref.\cite{kim}.

It should be  noticed that, as a reflection of the
growing complexity of the critical singularity structure,
  the uncertainty of our numerical 
results is  very low for $N=1$ and  relatively modest for
$N=0$, but it is much larger in the $N=2$ case. 
 However our new  estimates appear 
to be generally    more accurate than previous ones.
  The corresponding calculation for $N>2$ will be discussed 
in next section.

\section{ Analysis of the HT series for $N > 2$}

In this range of values of $N$ 
the general features of $N$-vector model 
 are expected to change qualitatively:   
reliable, although nonrigorous (and sometimes 
questioned \cite{seiler,casti}), 
RG calculations at low temperature \cite{polyakov}
indicate that the  model
is asymptotically free, namely that it becomes 
critical only at zero temperature.
The asymptotic behaviors of the  " second moment" 
correlation length, of  the susceptibility
 and of $ \chi_{4}$
as $\beta \rightarrow \infty$  are predicted to be

\begin{equation}
\xi^{as}(N,\beta) =c_{\xi}(N) 
\big({{-\beta} \over{b_0(N)}}\big)^{{b_1(N)} \over {b_0(N)^2}} 
exp[{{-\beta} \over{b_0(N)}}]  
\Big[ 1 + {{H_1(N)} \over{\beta}} + {{H_2(N)} \over{\beta^2}}+
O({{1} \over{\beta^3}}) \Big] \label{xias}\end{equation}

\begin{equation}
\chi^{as}(N,\beta) =c_{\chi}(N) 
\big({{-\beta} \over{b_0(N)}}\big)^{{{2 b_1(N)} \over {b_0(N)^2}} 
+{{g_0(N)} \over{b_0(N)}}}
exp[{{-2\beta} \over{b_0(N)} }] 
\Big[ 1+ {{K_1(N)} \over{\beta}} + {{K_2(N)} \over{\beta^2}}
+ O({{1} \over{\beta^3}}) \Big] \label{chias}\end{equation}

\begin{equation}
\chi_{4}^{as}(N,\beta) =-c_{4}(N) 
\big({{-\beta} \over{b_0(N)}}\big)^{{{6 b_1(N)} \over {b_0(N)^2}} 
+{{2g_0(N)} \over{b_0(N)}}}
exp[{{-6\beta} \over{b_0(N)} }] 
\Big[ 1 
+ O({{1} \over{\beta}}) \Big] \label{chi4as}\end{equation}

where 

\begin{equation}
b_0(N) \equiv {-{(N-2)}\over{2 \pi}},\;\;\;\;\;
     b_1(N)\equiv {-{(N-2)}\over{(2 \pi)^2}}, \;\;\;\; \;
   g_0(N)\equiv {{N-1}\over{2 \pi}}
\label{betagamma}\end{equation}

  are the first (renormalization scheme independent) low temperature 
perturbation 
expansion coefficients of the RG beta and gamma functions
\cite{polyakov} and $c_{\xi}(N)$,  $c_{\chi}(N)$ and   
$c_{4}(N)$  are universal quantities 
which clearly cannot be computed
 in (low temperature) perturbation theory. The (nonuniversal)
constants  $H_1$, $H_2$, $K_1$ and $K_2$   
appearing in (\ref{xias}) and (\ref{chias})
can be calculated in low temperature perturbation theory,
 and, on 
the square lattice \cite{falcionitreves,carapeli},  they come out
rather   small 
but not completely negligible in the range of values of $\beta$
 in which we shall be able to compute reliably $\chi$ and $\xi$ .
Numerical estimates for $H_1$, $H_2$, $K_1$ and $K_2$   
can be found in Ref.\cite{carapeli} and for brevity are 
not reported here, although we use them in the calculations. 
Unfortunately, the analogous
$O({{1} \over{\beta}})$ corrections have 
not yet been computed for $\chi_4$.  
As a  consequence of (\ref{xias}), (\ref{chias}) and (\ref{chi4as}), 
  for large $\beta$,

\begin{equation}
\hat g_r^{as}(N,\beta) =\hat g_r(N) [1+  O({{1} \over{\beta}})] =
\frac {\hat c_4(N)} {c_{\xi}(N)^2 c_{\chi}(N)^2} 
[1+  O({{1} \over{\beta}})].
\label{gierre} \end{equation}
 
where $\hat c_4(N) \equiv Nc_4(N)$.

Let us notice that the asymptotic formula (\ref{xias}) is 
valid both for the  "second moment" correlation length
$\xi(N,\beta)$ (which here is employed exclusively), and for
"exponential correlation 
length" $\xi^{exp}(N,\beta)$ 
but, a priori, with different 
multiplicative constants $c_{\xi}(N)$ 
 and $c_{\xi}^{exp}(N)$, respectively. 
However, it has been repeatedly noticed that 
$\xi^{exp}$ and $\xi$  are numerically very close in the critical 
region\cite{edwards}.
This fact is confirmed by a recent 
analytic calculation for large $N$ of the (universal) ratio 
$c_{\xi}^{exp}(N)/c_{\xi}(N)$ 
 giving the result\cite{cara}

\begin{equation}
c_{\xi}^{exp}(N)/c_{\xi}(N)= 1 + 0.003225/N + O(1/N^2)
\label{unosuenne}
\end{equation}

 Moreover,  for $N=3$, this ratio  
has been measured\cite{meyer}
 by a high precision MC method at $\beta=1.7$ and $1.8$,  
fully confirming the
quantitative reliability of the $1/N$ expansion 
 (\ref{unosuenne}) down to very low values of $N$ .
 Therefore, with very good approximation, 
we are justified in simply identifying $c_{\xi}^{exp}(N)$ 
 and $c_{\xi}(N)$ 
 even for small $N$.

These results are of direct interest here because 
the coefficient $c_{\xi}^{exp}(N)$
can be  computed exactly \cite{hasenf}  
by the thermodynamical Bethe Ansatz and its value, 
 with our normalization conventions,  is

\begin{equation}
c_{\xi}^{exp}(N)=\frac{\Gamma(1+\frac {1} {N-2})} {\sqrt{32}} 
exp\Big[ -\frac {\pi/2 + log(8/e))} {N-2} \Big] \simeq c_{\xi}(N) 
\label{bethe}\end{equation}

Let us now turn to series analysis.
With sufficiently long series, like those analyzed here, 
even the simple plot of the HT coefficients of $\chi$, 
 versus their order, 
 reported in Fig.3, 
is suggestive enough  that  
the series for $N > 2 $ must behave quite differently 
from those  for $N \leq 2$. 
 Notice that in the plot the coefficients 
have been conveniently normalized, 
 as indicated in the figure caption, in order 
to make the behaviors for different values of $N$ easily comparable.
 In the $N \leq 2$ case the 
coefficients  are positive (and monotonically
increasing) in agreement with the 
fact that the nearest singularity
is located on the real positive 
$\beta$ axis (and that the antiferromagnetic 
singularity located at $-\beta_c$ is much weaker).
On the contrary, in the  $N > 2$ case the coefficients 
do not remain positive as their 
order grows and they display what can be 
safely interpreted as the onset of an oscillatory trend. 
This feature is related 
  via Darboux theorem \cite{olver} to   the fact  
that, for $N > 2$,  the nearest singularities of $\chi$ 
in the complex $\beta$ plane become unphysical, as we
 have first pointed out some time ago\cite{bcm90}.
This applies also to $\mu_{2}$, as well as to $\chi_4$ 
with the only difference that for these quantities 
the oscillating behavior of the 
expansion coefficients should set in at higher orders.
Our interpretation of these general features is impressively 
confirmed by a study\cite{bcmo} of the spherical model 
 which, in spite of a priori legitimate 
mistrust about exchanging the large 
$\beta$ and the large $N$ limits, turns out to 
be a completely reliable guide to the 
qualitative behavior of the $N-$vector 
model even for not too large $N \geq 3$, as it has been
 also  argued some time ago 
 \cite{bcm90,brout}.

As already mentioned above, an arbitrarily large 
number of HT expansion coefficients for $\chi^{(s)}$, 
 $\mu_{2}^{(s)}$  and $\chi_{4}^{(s)}$ of the spherical model 
can be easily computed, for any lattice. 
In the case of the square lattice 
they exhibit regular cyclic 
alternations in sign of period 8
 related 
to the symmetric quartet structure of 
the nearest unphysical singularities
in the complex $\beta$ plane. We  have accurately mapped out  
in Ref.\cite{bcmo} the whole  set of singularities, all of which
 are square root branch points.
 In the vicinity of $\beta=\infty$,  this set 
 has the characteristic 
structure dictated by asymptotic freedom (which was first  
discussed for the case
of QCD in Ref.\cite{thooft}), the 
analiticity domain of  $\chi^{(s)}$, 
being a wedge with zero opening
 angle  which contains the real $\beta$ axis. It is quite likely 
that these features of the spherical model 
persist also down to all finite $N \geq 3$ 
\cite{nota}, although a complete
study of this question is presently infeasible.

The transition from the $N < 2$   regime 
characterized by a conventional power law critical point,
to the $N > 2$ asymptotically free
 regime characterized by unphysical 
singularities  can  be closely followed
  by locating the position
of the nearest singularity $\beta_n$  
in the first quadrant of the complex $\beta$ plane as a function of 
$N$.  For convenience in the graphical representation
 of the results, we shall use in what follows  the  scaled  variable
$\tilde \beta \equiv \beta/N$ and  plot the estimates of 
$\tilde \beta_n$ versus $x \equiv 1-1/N$, rather than versus
$N$. The trajectory of the singular point $\tilde \beta_n$ 
as a function of $x$ in the complex 
$\tilde \beta$ plane can be traced out  as described 
in Ref.\cite{bcm90} either by using 
  PAs to locate the nearest singularity of the log derivative of
 $\chi$ or by directly computing DA's of $\chi$.
 The results of both procedures agree perfectly  
within the numerical uncertainties.

  In Fig.4 
 we have plotted the real and the imaginary part
of  $\tilde \beta_n$ 
as  functions of $x$ in the range $0 \leq x \leq 1$.
For $ -2 \leq N \leq \bar N \protect \lesssim 2.2$, the singularity 
$\tilde \beta_n$ is still a real
critical point, but for $N > \bar N$ 
it splits into a pair of complex conjugate 
singularities which move into the complex plane and,
 as $N \rightarrow \infty$,  tend to the limiting
points $\tilde \beta_{\pm} \simeq 0.32162(1 \pm i)$. 
In particular for
$N=3$ the nearest pair is 
located at $\tilde\beta_n=0.58(5) \pm 0.14(5)i$, 
 while for $N=4$ we have $\tilde\beta_n=0.55(4) \pm 0.22(5)i $,
for $N=6$ we have $\tilde\beta_n=0.50(2) \pm 0.28(2)i$ and for
$N=10$ we have $\tilde\beta_n=0.44(2) \pm 0.31(1)i$.
Although in general they may be weaker,
 the corresponding antiferromagnetic singularities 
will follow trajectories 
symmetrical with respect to the $Im (\tilde \beta)$ axis
 so that, for all $N > \bar N$  the set of 
the nearest singularities will 
form a  quartet with the same symmetry.

 It is certainly conceivable\cite{seiler} that, in contrast with the 
 perturbative RG predictions, 
 when the nearest singularities become
 complex,  a further real critical singularity might appear so that, 
even for $N \geq 3 $, it would 
 be still possible to relate
 the steep dependence on $\beta$ of $\chi$,  $\xi$ and $\chi_4$
 to a conventional finite temperature phase transition, 
 but  we have not been able to find any numerical indication of 
such a possibility for 
not too large $\tilde\beta$.  
More precisely neither Dlog PA's nor DA's 
exhibit any real and numerically 
stable singularity in their range of sensitivity.
Another argument against the existence  of critical points 
 for finite values of $\beta$ comes from the
 observation illustrated below, that,  by a simple  procedure,
  the high temperature behavior  of $\xi$, $\chi$  and $\chi_4$
can be smoothly extrapolated onto their low
temperature behavior  (\ref{xias}), (\ref{chias})  and (\ref{chi4as})
as predicted by the perturbative RG.
This is feasible for any $N$, although  the procedure
 is numerically  very accurate 
 only for $N > 3$, since, for $N=3$, $Im(\tilde \beta_n)$
is  small and therefore 
the behavior of $\chi$ or $\xi$ on the real $\tilde\beta$ axis is 
 more strongly
 perturbed  in the vicinity of $Re(\tilde \beta_n)$.
 Both  the often reported 
failure in  observing asymptotic scaling
by MC simulations of the $N=3$ model 
at moderate values of $\tilde \beta$
and  the better successes for larger values of $N$ find 
 a completely plausible explanation in this picture.
Of course, the results of our extrapolation scheme 
would be difficult to explain if the high temperature 
region were separated
 by a critical point  from the low temperature region. 

Let us now  describe an approximation scheme which enables us 
to estimate   low temperature perturbative
parameters such as  $b_0(N)$, as well as nonperturbative 
parameters like $c_{\xi}(N) $, $c_{\chi}(N)$ etc. 
entering into the asymptotic formulas
 (\ref{xias}), (\ref{chias}) and (\ref{chi4as}), 
in terms of our HT series.
 Since  
$\xi$,  $\chi$ etc. are exponentially fast 
varying quantities at large values of  $\beta$, 
neither  PA's nor DA's are well suited  for a straightforward 
extrapolation of the HT series from small 
to (relatively) large $\beta$ values. 
We should rather work with quantities
which vary slowly enough to be well represented by PA's or DA's.
 Let us observe that,  
if $\chi$ has the asymptotic behavior (\ref{chias}),  
then for large enough  $\tilde \beta$
\FL
\begin {eqnarray}
B_{\chi}(N,\tilde \beta) 
\equiv {{1}\over{2}} Dln[\chi (N, \tilde \beta)] 
+ {{N+1}\over {2(N-2)\tilde\beta}}- {{1}\over{2}} 
Dln[1+ {{K_1} \over{N \tilde \beta}} 
+ {{K_2} \over{N \tilde \beta^2}}]  
 \simeq -N/ b_0(N) 
+O({\frac {1} {N^4 \tilde\beta^4}}).
 \end{eqnarray}

 The log 
derivative of $\chi$, which is a slowly varying 
quantity, can be approximated 
by near diagonal PA's and 
then we can reliably extrapolate the 
quantity  $B_{\chi}(N,\tilde \beta)$. 
In practice, due  to the
finite extension of our series and 
to the intricate analytic structure of $\chi$, we do not expect
that this is a  good approximation for  large $\tilde\beta$
and we  rather make the reasonable  (and successful) assumption that  
 the  $\tilde\beta$ independent parameter $b_0(N)$
 is best approximated by evaluating   
$B_{\chi}(N,\tilde \beta)$   at 
some finite real value 
$\tilde\beta = \tilde\beta_s$
 where it is stationary or it shows the 
slowest variation when $\tilde\beta$ is varied.
 Consistency of this approximation 
scheme requires that, as the number
of HT coefficients used in the calculation 
is increased, the stationary value 
$B_{\chi}(N,\tilde \beta_s)$ stabilizes 
and  that $\tilde \beta_s \rightarrow \infty$.
It can be checked that this actually happens 
in the $N=\infty$ case  in which arbitrarily long
HT expansions can be studied and also that  our approximation scheme 
converges rapidly to the expected result.
 
A further check of the correctness of our procedures comes 
from the obvious remark that similar estimates of the same parameter 
$b_0(N)$ should be obtained 
starting either with the correlation length 
 $\xi$ and computing the quantity
\FL
\begin{eqnarray} 
B_{\xi}(N,\tilde \beta) \equiv {{1}\over{2}} 
Dln[\xi (N, \tilde \beta)^2/\tilde \beta] 
+ {{N}\over {2(N-2)\tilde\beta}} 
-Dln[1+ {{H_1} \over{N \tilde \beta}} 
+ {{H_2} \over{N^2 \tilde \beta^2}}] 
\simeq 
 -N/ b_0(N) 
+O({\frac {1} {N^4 \tilde\beta^4}}).
\label{bizeroxi} 
\end{eqnarray}

or starting with $\chi_{4}$ and computing the quantity 
\FL
\begin {equation} 
B_{4}(N,\tilde \beta) \equiv {{1}\over{6}} 
Dln[\chi_4 (N, \tilde \beta)] 
+ {{(N+2)}\over {3(N-2)\tilde\beta}} \simeq -N/ b_0(N) 
+O({\frac {1} {N^2 \tilde\beta^2}}).
\label{bizerochi4} 
\end{equation}

In Fig.5
  we have plotted 
$B_{\chi}(N,\tilde \beta)$ versus $\tilde \beta$  
for various values of $N$,  
 in order to show that  a  stationary 
point  $\tilde\beta_s(N) $ actually exists
 around $\tilde\beta \simeq 0.55$ for any $N$, and that 
the  size  of the neighbourhood of $\tilde\beta_s(N)$
where $B_{\xi}(N,\tilde \beta)$ varies  
slowly with $\tilde \beta$,  grows with $N$. 
Notice that  
$ Re (\tilde \beta_s(N)) \protect \lesssim 0.55$ 
for $N>4$ and therefore on the border
of the convergence region of the series or slightly outside it.

In Fig.6 we have plotted versus $N$ 
our numerical estimates of $b_0(N)$ from the quantities 
$B_{\chi}(N,\tilde \beta_s(N)),B_{\xi}(N,\tilde \beta_s(N))$ and 
$B_{4}(N,\tilde \beta_s(N))$  
 and have compared them to the expected value (\ref{betagamma}). 
Each point represents the average of the near diagonal PA's 
using at least 14 series coefficients 
for the quantities  $B_{\chi}(N,\tilde \beta_s)$ and 
$B_{\xi}(N,\tilde \beta_s)$,  and at least 10 
coefficients for $B_{4}(N,\tilde \beta_s)$. 
We have reported  relative errors of $5\%$  
 suggestive both of the scatter of 
the estimates obtained by the various PA's and of the 
systematic uncertainties of our extrapolation procedure.

In conclusion, it appears that 
from our high temperature data  for $\xi$,  $\chi$
and $\chi_4$, 
 we have  been able to extract  completely 
consistent and correct estimates of the low temperature 
perturbation parameter $b_0(N)$
which characterizes the 
exponential asymptotic growth of these quantities,
and  in general that the deviation from the 
expected value (\ref{betagamma}) of $b_0(N)$
is never larger than $5\%$ over a wide range of values of $N$.

In quite a similar way, assuming that $b_0(N)$ 
is given by (\ref{betagamma}), 
we can estimate the exponents of the power law
prefactors in  (\ref{xias}), (\ref{chias}) and (\ref{chi4as}). 
As it must be expected, 
the errors in this computation are somewhat higher, but
 they do not exceed $20-30\%$.

Let us now show that
by a similar  approximation procedure we 
can also estimate the  constant $c_{\xi}(N)$.
 We have simply to compute the 
HT series of the slowly varying quantity 

\begin{eqnarray}
C_{\xi}(N,\tilde\beta) \equiv \xi^2(N,\tilde\beta)
exp[{{2N \tilde\beta} \over{b_0(N)}}] \simeq \\ \nonumber
c_{\xi}(N)^2 \big({{-N \tilde \beta} \over{b_0(N)}}\big)^{{-2} 
\over {N-2}}\Big (1 + {{H_1} \over{N \tilde \beta}} + {{H_2} 
\over{N^2 \tilde \beta^2}}+ O(\frac {1} {N^3\tilde \beta^3})\Big)
 \end{eqnarray}

obtained by  dividing out the exponential factor 
in the asymptotic behavior (\ref{xias}) of $\xi^2$.

 We then  form  near  diagonal PA's to $C_{\xi}(N,\tilde\beta)$
and use them to evaluate the quantity 

\begin {equation} 
C_{\xi}(N,\tilde\beta)({{-N \tilde \beta} 
\over{b_0(N)}}\big)^{{2} \over {N-2}}\Big 
(1 + {{H_1} \over{N \tilde \beta}} 
+ {{H_2} \over{N^2 \tilde \beta^2}} \Big)^{-2} 
\label{cxi} \end{equation}

 at the 
value $\tilde\beta_s$  where it is stationary.
In analogy with the  previous computation this is taken 
to be the best approximation of $c_{\xi}(N)^2$. It 
is observed that also in this 
case the stationary values occur for $\tilde\beta \simeq 0.5$.

 Similarly, we  can estimate $c_{\chi}(N)$ 
 by studying the HT series for the quantity

\begin{eqnarray}
C_{\chi}(N,\tilde\beta) \equiv \chi(N,\tilde\beta)
 exp[{{2N \tilde\beta} \over{b_0(N)} }] \simeq \\  \nonumber
c_{\chi}(N)\big({{-N\tilde \beta} 
\over{b_0(N)}}\big)^{{-N-1}\over {N-2}} 
\Big(1  +{{K_1} \over{N \tilde \beta}} 
+ {{K_2} \over{N^2 \tilde \beta^2}}+ 
 O(\frac {1} {N^3\tilde \beta^3})\Big).
 \end{eqnarray}

Unfortunately no exact formula is known for $c_{\chi}(N)$ , but 
 we can compare our numerical estimates to the $1/N$ expansion 
 through $O(1/N)$  of $c_{\chi}(N)$ which has been  computed 
in Ref. \cite{carapeli}

\begin{equation}
c_{\chi}(N)  = \frac {\pi} {16} [ 1 - \frac{4.267} {N} + O(1/N^2)]
\end{equation}

 or to
an analytic formula  recently guessed in Ref.\cite{caralat95}
 with no other theoretical justification than a
 formal analogy with the exact 
formula (\ref{bethe}) for $c_{\xi}(N)$.

An estimate of $c_4(N)$ could be obtained   
starting with the series for 

\begin{equation}
C_{4}(N,\tilde\beta) \equiv \chi_4(N,\tilde\beta)
 exp[{{6 N \tilde \beta} \over{b_0(N)} }]  \simeq 
c_{4}(N)\big({{-N\tilde\beta} 
\over{b_0(N)}}\big)^{{-2 (N+2)}\over {N-2}}
(1   + O(\frac {1} {N\tilde \beta})).
\label{cchi4} \end{equation}

However the $O(\frac {1} {N\tilde \beta})$ 
corrections are not known, 
and moreover the  $1/N$ expansion of $c_{4}(N)$ 
which has been computed in Ref.\cite{campogr} 

\begin{equation}
c_4(N)= \frac {\pi^3} {1024N} [1.- 15.6/N + O(1/N^2)]
\end{equation}

  it is practically useless, except for very large $N$, 
since  the subleading term is quite large. 
Therefore we do not report our estimates for $c_4(N)$.

 By the same method we have also directly estimated the universal 
quantity $c_r(N) \equiv 
(\frac {2\pi N} {N-2} )^{\frac {N-1} {N-2}} 
c_{\xi}^2(N)/c_{\chi}(N) $ 
which appears in the 
asymptotic expression of the ratio 

\begin{equation}
\frac {\xi^2} {\chi} \simeq
c_r(N)\tilde\beta^{\frac {N-1} {N-2}}
(1 + {{H_1} \over{N \tilde \beta}} 
+ {{H_2} \over{N^2 \tilde \beta^2}}+ 
 O(\frac {1} {N^3\tilde \beta^3}))^2
(1  +{{K_1} \over{N \tilde \beta}} 
+ {{K_2} \over{N^2 \tilde \beta^2}})^{-1}.
\end{equation}

 and have compared it to the $1/N$
expansion 
\begin{equation}
c_r(N)= 1+ 1.955/N + O(1/N^2). 
\label{cr} \end{equation}

Let us notice that, for large $N$, the
unphysical singularities of 
$\xi^2$ and $\chi$ tend to cancel in the ratio
and that the $1/N$ correction  in (\ref{cr}) is not very large.

We have reported  our numerical 
estimates for $c_{\xi}(N)$, $c_{\chi}(N)$  
and $c_{r}(N)$ 
in Table 1 and in Fig.7   where 
 they are compared  to the exact or conjectured formulas
and  to their $1/N$ 
expansions. Like in the previous Fig.6, 
the error bars we have attached 
to our data points are fairly 
subjective in that they include a  "statistical" contribution
 (describing the spread of the estimates from various approximants)
 which is not large in general, while the
main part of the  uncertainty comes from our  
estimate of the  systematic error. As it appears from the 
Fig.7 and from Table 1, 
 our central estimates for $c_{\xi}(N)$ and the exact  
formula agree within $1-2\%$ 
 on the whole range of $N$  except for the lowest values of $N$. 
It should be observed that
we have not our reported estimates  for $N=3$ since in this case 
the nearby unphysical 
singularities have a very small imaginary part 
$Im(\beta_n)$ and there is a large spread in the stationary values 
of $C_{\xi}$  and $C_{\chi}$. This 
makes difficult to estimate unambiguously 
the values of $c_{\xi}$  and $c_{\chi}$. 
However, if we shift to only slightly 
larger values of $N$, such as $N= 3.5$, then $Im(\beta_n)$ is
already sufficiently large 
for our procedure to work appropriately and 
we can estimate  $c_{\xi}(3.5)=0.028(8)$ 
to be compared to the exact value 
$c^{exp}_{\xi}(3.5)=0.0273$  and, 
similarly, $c_{\chi}(3.5)=0.021(2)$, 
 while the conjectured formula gives $c^{exact}_{\chi}(3.5)=0.0266$.
In both cases the discrepancy is less than $20\%$.
 At $N=4$ 
 the exact value is $c^{exact}_{\xi}(4)=0.0416 $ and  we 
find $c_{\xi}(4)=0.039(1)$, which is off only by $6\%$. 
For larger $N$ the agreement 
is much closer as it is shown in Table 1. 
Our estimates 
 also agree  well with the conjectured 
exact formulas  for $c_{\chi}(N)$ and $c_r(N)$.
 The discrepancy 
from these formulas or from their 
 $1/N$ expansions does not exceed  $5-10\%$ for small values of $N$ 
but it gets significantly smaller
already for moderately  large $N$.
All  numerical results are collected in Table 1.  
We believe that both the failure to reproduce 
 accurately the $N=3$ values of the parameters and the other
 general features of our approximations 
should not be surprising if we take into account 
the analytic structure in the $\tilde\beta$ complex plane
  of the quantities to be extrapolated  and 
we consider  that  our 
computational method is the simplest  and most 
direct possible  and  
 also that we are still working 
at  moderate values of $\tilde \beta$
 where, for small values of $N$, 
the correlation length is not very large.  
Our approximation procedures should not 
however be suspected to be "ad hoc", 
 since they were proposed  and  the first 
results \cite{bcm90} were published 
 before either  the exact formula(\ref{bethe}) and   
the $1/N$ expansions  became known. 

We should also   at this point recall 
  our remark \cite{bcmo} (resumed  and 
applied to fourteen term series 
in \cite{honte}) that the precision of 
these estimates  might  be significantly 
improved by performing a conformal 
transformation of the complex $\tilde \beta$ plane in order to
remove at least the quartet of the nearest unphysical singularities
 before applying our approximation procedure. 
We think, however, that the success of 
our  straightforward treatment
 over a wide range of values of $N$ cannot be accidental and that 
simply getting a higher level of accuracy  could hardly be 
more convincing  of the validity of the RG
picture of scaling and indicative 
of the purely numerical origin of the 
 discrepancies for the lowest values of $N$. Therefore we
 shall not pursue  here our old  suggestion.

Finally, we can also estimate  $\hat g_r(N)$  for $N \geq 3$,  
by forming PAs to the series expansion of $1/{\hat g_r(N,\beta)}$
 and evaluating them at their stationary points. 
We have reported   in Fig.2 our  estimates 
and have included for comparison  the field theoretic 
estimate\cite{falcioni} for $N=3$,
and  MC estimates \cite{falcioni,basu,kim}  and other HT 
series estimates \cite{campogr}  for $N=3$ and  $N=4$.
It should also be noticed  that our results are entirely consistent
with the large $N$ limit, in which we have 
\begin {equation} 
\hat g_r(N) =  {8\pi}[ 1 - 0.602033/N +O(1/N^2)].
\label{grn} \end{equation}

The  $1/N$ correction
 has  been computed recently\cite{campogr}.
Also   the accuracy of  this calculation is satisfactory and
 the maximum  error, for $N \geq 3$,   
can be rated  not to exceed  $5\%$.
Results and conclusions in complete 
agreement with ours are reached in 
the somewhat different analysis of the HT series 
 presented in Ref.\cite{campo2}.

\section{ Conclusions}

We have presented our estimates of the low temperature quantities 
$b_0(N)$, $c_{\xi}(N)$ etc. defined by  
 (\ref{xias}), (\ref{chias}) and (\ref{chi4as}), 
 obtained by a procedure 
 which can essentially be seen as a simple 
improvement of the "matching method" 
introduced long ago in Ref.\cite{nelfi} 
and since used 
several times with more or less unconvincing results, due 
either to inadequate
 implementation and/or to incorrect supplementary assumptions.
The initial paper \cite{nelfi}  is an example of the former defect: 
 the low temperature behavior was 
inadequately accounted for by one loop 
perturbation expansion and, on the HT side, too
 short series were used  resulting into an 
unreliable  matching.
On the other hand  Ref.\cite{parisi} 
is an example of both shortcomings since the use of  HT
series (at that time extending to 
ten terms only)   was supplemented 
with the (now  appearing obviously incorrect) 
conjecture that $\chi$ and $\mu_2$ have all positive
  HT coefficients.
Indeed even if we made the weaker assumption that there 
are at most finitely many
negative expansion coefficients this would clearly imply that the
nearest singularity of (for example) 
$\chi$ is located on the real positive 
$\beta$ axis. If also  asymptotic  freedom holds, 
then  $\chi$  should be a regular analytic function 
in the whole finite complex $\beta$ 
plane, contrary to the numerical evidence 
presented in the previous section.

 We have tried to avoid the shortcomings of 
the previous approaches by 
 the simplest possible treatment of sufficiently long HT series and 
 by excluding unwarranted supplementary assumptions.

A  brief 
review of  some earlier references which are 
 closely related to our
analysis already appears in \cite{bcm90}.  Here we shall 
mention only some later studies and  address the reader to 
Ref.\cite{bcm90} for a long (but surely  still incomplete) 
list of the  abundant prior literature. 

It is worthwhile to recall that  recently, in the $N=3$
case, a new method for extrapolating 
finite volume MC data to infinite 
volume\cite{cara}   has been used to test the onset of the 
asymptotic behavior (\ref{xias})
by obtaining the second moment correlation length 
$\xi$ up to values as large as $10^5$ lattice units and agreement 
with (\ref{xias}) has been 
found within $4\%$ at $\tilde\beta=1.$ where 
$\xi \simeq 10^5$.
 New MC data  are now available\cite{caralat95} also 
for the susceptibility which yield
$c^{MC}_{\chi}(3)= 0.0146(10)$, $c^{MC}_{\chi}(4)= 0.0383(10)$ and 
$c^{MC}_{\chi}(8)= 0.103(2)$ in 
very good agreement with our estimates 
$c_{\chi}(4)= 0.0344(3)$ and 
$c_{\chi}(8)= 0.1037(4)$ as well 
as with the conjectured exact results 
$c^{exact}_{\chi}(4)= 0.0385$ and 
$c^{exact}_{\chi}(8)= 0.1027$.
 Analogous results for the correlation length in the $N=3$ case 
had been presented also in a 
computation\cite{kimxi} extending  to  $\xi \simeq 15000$.
In an earlier  high precision multigrid MC study  
devoted to the $N=4$ model on 
lattices of size up to $256^2$ \cite{edwards},
 the asymptotic behavior of 
$\xi$ and $\chi$ had been found to be perfectly 
compatible with  (\ref{xias})  and (\ref{chias})
 and the quantities $c_\xi(4)$ and $c_\chi(4)$ had been 
estimated to be 
$c_\xi(4)=0.0342(20)$ and $c_\chi(4)=0.0329(16)$.
Moreover in that study the possibility
of an ordinary critical point 
 $\tilde\beta_c \leq 1.25$ was excluded, and it was stressed
that the data could be compatible with a 
value $\tilde\beta_c \geq 1.25$ only 
assuming implausibly large values for the critical exponents. 
( A far away power singularity with a large exponent 
is  likely to be merely an effective 
representation of an exponential 
behavior.)
Also the  MC single cluster  simulation of Ref.\cite{wolff} for 
the $N=4$ and $N=8$ models
  gave good support to the asymptotic formulas (\ref{chias}) 
and (\ref{xias}) and produced 
 estimates for $c_\xi$  completely consistent with (\ref{bethe}). 

Finally, on the side of the analytic approaches,  we should   
mention the study of the scaling behavior in Refs.
\cite{fly}, whose results include a computation
of the leading term of the $1/N$ expansion of $c_\xi$, 
in complete agreement with the exact result (\ref{bethe}), 
 and of the same expansion for $c_\chi$. 

In conclusion, we can summarize our main results as follows:

a) By this and previous work \cite{bcm90} we have  shown that 
our general $N$ HT series are a useful tool also  for obtaining high 
precision estimates of critical 
parameters in somewhat unconventional
contexts, giving further support to qualitative and quantitative
results obtained by entirely different approximation methods. 

b) In the  $ -2 \leq  N < 2$ vector models case we  have confirmed,
 with high accuracy, 
the explicit formulas obtained by  (semirigorous) model solving, 
for the critical exponents $\gamma(N)$, $\nu(N)$ and $\Delta(N)$.
We have also computed the "dimensionless renormalized 
four point coupling  constant" $\hat g_r(N)$  for $N=0,1,2$
in complete agreement with other estimates, 
but with higher accuracy.

c) For the $N > 3 $ vector models, we can somehow extrapolate the 
HT series to the border of (or  beyond)
their region of convergence reliably enough
 to  reconstruct the quantitative 
features of low temperature behavior 
and we can obtain  a set 
of (hardly accidental)  consistency checks  
with the predictions of  the  perturbative RG, 
of exact solutions and of $1/N$ 
expansions  with an accuracy practically  uniform 
with respect to $N$. 
As shown by Table 1, our estimates of the parameter $c_{\xi}(N)$ 
agree well with the exact calculation by the Bethe 
Ansatz (under the assumption (\ref{bethe})). On the other hand
our estimates for $c_{\chi}(N)$, $c_{r}(N)$ and $\hat g_r(N)$ 
are completely consistent with their $1/N$ expansions.

Of course we must say that, strictly speaking, 
purely  numerical computations 
cannot validate the RG
 predictions: only complete proofs can settle the question, 
but they are still to come. Therefore, 
in principle,  further discussion of this subject 
 may still be considered
 healthy and welcome 
as long as it may  stimulate 
either to design a rigorous justification 
 of the generally accepted RG picture or to 
produce viable mechanisms  for evading the 
expected asymptotic freedom 
regime while respecting the now established  heuristic evidence.  
We believe, however, that the 
continuing accumulation of unambiguous, 
consistent and increasingly accurate numerical support for the 
RG predictions from  a variety of 
independent  approaches 
leaves little if any space for alternative pictures.

\acknowledgements
This work has been partially supported by MURST.

\appendix

\section{The susceptibility}

The HT coefficients of the susceptibility 
$\chi(N,\beta) =   1+ \sum_{r=1}^\infty a_r(N) \beta^r $ are

$a_1(N)= 4/N$

$a_2(N)= 12/N^2$

$a_3(N)=(72 + 32 N)/{(N^3 (2 + N))}$

$a_4(N)=  (200 + 76 N)/{(N^4 (2 + N))}$

$a_5(N)= 8 (284 + 147 N + 20 N^2)/{(N^5 (2 + N)(4 + N))}$

$a_6(N)= 16(780+719N + 201N^2 + 19N^3)/{(N^6 (2 + N)^2 (4 + N))}$

For the  coefficients  which follow it is typographically more 
convenient to set 
$a_r(N)=P_r(N)/ {Q_r(N)}$  and to tabulate separately the numerator
polynomial $P_r(N)$ and the denominator polynomial  $Q_r(N)$.

\FL

$P_{7}(N)=16(26064 + 38076 N + 20742 N^2 
+ 5280 N^3 + 655 N^4 + 32 N^5)$

$Q_{7}(N)=N^7 (2 + N)^3 (4 + N)(6 + N)$

\FL

$P_8(N)= 4 (283968 + 383568 N + 186912 N^2 
+ 41000 N^3 + 4392 N^4 + 187 N^5)$

$Q_8(N)=   N^8 (2 + N)^3 (4 + N) (6 + N)$

\FL
$P_{9}(N)=8(3123456+4186336 N+2087128 N^2+492220 N^3+
62386 N^4+4161 N^5+116 N^6)$

$Q_{9}(N)= N^9 (2 + N)^3 (4 + N) (6 + N) (8 + N)$

\FL
$P_{10}(N)=  16 (33868800 + 66758016 N + 53214272 N^2 
+ 22126648 N^3 
+ 5211372 N^4 + \\ 719330 N^5 + 58789 N^6 +
 2684 N^7 + 55 N^8)$

$Q_{10}(N)=   N^{10} (2 + N)^4 (4 + N)^2 (6 + N) (8 + N)$

\FL
$P_{11}(N)= 32 (3695370240 + 9913385984 N + 11437289216 N^2 
+7427564992 N^3 + \\
2989987696 N^4 + 776848144 N^5 + 132130072 N^6 + 14693596 N^7 +  
1052911 N^8 + \\
 46923 N^9 + 1225 N^{10} + 16 N^{11})$

$Q_{11}(N)=   N^{11} (2 + N)^5 (4 + N)^3 (6 + N) (8 + N) (10 + N)$

\FL
$P_{12}(N)= 16 (4990955520 + 11511967232 N + 
10992991488 N^2 +5609888352 N^3 + \\
  1649559472 N^4 + 281912408 N^5 + 27080244 N^6 
+1334568 N^7 +  22368 N^8  
 - 199 N^9 + 5 N^{10})$

$Q_{12}(N)=   N^{12} (2 + N)^5 (4 + N)^2 (6 + N) (8 + N) (10 + N)$

\FL
$P_{13}(N) = 64 (162478080000 + 406158981120 N + 431982472192 N^2 + 
254291324928 N^3 
+ \\  90288340864 N^4 + 19721001832 N^5 + 2561904944 N^6 +
170376718 N^7 +  1211742 N^8 \\ - 616479 N^9 - 37625 N^{10} -
 635 N^{11} + 4 N^{12})$

$Q_{13}(N)= N^{13} (2 + N)^5 (4 + N)^3 (6 + N) 
(8 + N) (10 + N) (12 + N)$

\FL
$P_{14}(N)=16(21007560867840 + 63770201063424 N 
+ 84400316350464 N^2 +  \\
 63787725946880 N^3 + 30245054013440 N^4 + 9275137432448 N^5 +  \\
 1810742519232 N^6 + 204284290016 N^7 + 7651128688 N^8 - 
 1135009056 N^9 - 177417600 N^{10}\\ 
- 9861216 N^{11}  - 161554 N^{12} + 
       5277 N^{13} + 172 N^{14})$

$Q_{14}(N) = N^{14} (2 + N)^6 (4 + N)^3 
(6 + N)^2 (8 + N) (10 + N) (12 + N)$

\FL
$P_{15}(N)=16(9537349044142080+34641374485413888 N 
+56169030631292928 N^2 +\\ 
  53537028436525056 N^3 + 33216830381735936 N^4 
+ 14006542675035136 N^5 + \\
 4047829991104000 N^6 + 777531907925504 N^7 
+ 87227510881024 N^8 +
 1944102682560 N^9  \\- 1061882170400 N^{10} 
- 183809104832 N^{11} -
 14563826832 N^{12} - 515944376 N^{13} + \\   5830192 N^{14} + 
       1259012 N^{15} + 43647 N^{16} + 512 N^{17})$

$Q_{15}(N)= N^{15} (2 + N)^7 (4 + N)^3 (6 + N)^3 (8 + N) (10 + N) 
(12 + N) (14 + N)$

\FL
$P_{16}(N)=4(410123375487221760+1537129944780374016 N
+2583983411690471424 N^2+\\ 
 2566975595695570944 N^3 + 1669084283351334912 N^4 + 
       741114014711103488 N^5 +\\ 
225948162044579840 N^6 +  45385102417264640 N^7 
+4996850176026624 N^8  \\
 - 64804204496896 N^9 - 122658733213440 N^{10} 
- 20909429640960 N^{11} -
1752208241536 N^{12} -  \\  56642417728 N^{13} 
+ 3062606512 N^{14} +
 412508368 N^{15} + 	18713696 N^{16} 
+ \\  395328 N^{17} + 3083 N^{18})$

$Q_{16}(N)=   N^{16} (2 + N)^7 (4 + N)^4 (6 + N)^3 
(8 + N) (10 + N) (12 + N) (14 + N)$

\FL
$P_{17}(N)=  8 (35361815028050165760 + 138539666887258669056 N +  \\
  245436909326998437888 N^2 +  259375081142913859584 N^3 + 
  181396134616565809152 N^4 +  \\87793764370648399872 N^5 +
  29673202500166647808 N^6 + 6770762601709142016 N^7 +  \\
  893508862130341888 N^8 +  5229394076767232 N^9 -
  24934992828139008 N^{10}  \\- 5547918408527104 N^{11} - 
  625740097598720 N^{12} - 33521607263744 N^{13} +\\
 738107699392 N^{14} + 
 272358030048 N^{15} + 21867513640 N^{16} + 937447020 N^{17} +  \\
 22261658 N^{18} + 250495 N^{19} + 692 N^{20})$

$Q_{17}(N)=N^{17} (2 + N)^7 (4 + N)^5 (6 + N)^3 (8 + N) (10 + N) 
(12 + N) (14 + N) (16 + N)$

\FL
$P_{18}(N)= -16 (-758936838540424642560 - 3343934774878956158976 N -  \\
  6720650800795024883712 N^2 - 8141819950133007089664 N^3 - 
  6611686534391523180544 N^4  \\- 3777832155850593533952 N^5 - 
  1543669324445646061568 N^6 - 443919373830353158144 N^7 -  \\
  82653304109539049472 N^8 -  6333077582288386048 N^9 + 
  1419388196952978432 N^{10}  + \\ 578178922758906368 N^{11} + 
  99290612095487744 N^{12} + 9300735775467264 N^{13} +  \\
  256677768200576 N^{14} - 53852331942080 N^{15} 
- 8516631212960 N^{16} - 
  629479458104 N^{17}  \\- 26896421724 N^{18} - 607483694 N^{19} -
       2912825 N^{20} +  154080 N^{21} + 2313 N^{22})$

$Q_{18}(N)= N^{18} (2 + N)^8 (4 + N)^5 (6 + N)^3 (8 + N)^2 (10 + N) 
(12 + N) (14 + N)(16 + N)$

\FL
$P_{19}(N)= -32 (-293792962132985669222400 - 
 1452203904509587992084480 N -\\ 3305764874023051160715264 N^2 - 
4587498272216547279765504 N^3 \\ - 4326547244550747303444480 N^4 - 
2922303204727243671601152 N^5  \\- 1446836388063996470624256 N^6 - 
524786164553898279829504 N^7  \\- 134502578329442459254784 N^8 - 
 21104878727348421885952 N^9  \\- 411991601001072488448 N^{10} + 
 794587338452494176256 N^{11} + 242141294836583751680 N^{12} +  \\
 39373307992978213888 N^{13} + 3583854665917282560 N^{14} + 
 51879072941552128 N^{15}  \\- 36069375006840576 N^{16} - 
 5868286096676352 N^{17} - 508264525824336 N^{18}  \\- 
27200961065872 N^{19} -  811699909040 N^{20} - 3015005636 N^{21} +
793163459 N^{22} +\\ 32254806 N^{23}+  562185 N^{24} + 3824 N^{25})$

$Q_{19}(N)=   N^{19} (2 + N)^9 (4 + N)^5 (6 + N)^3 (8 + N)^3 (10 + N) 
(12 + N) (14 + N) (16 + N) (18 + N)$

\FL
$P_{20}(N)= 25184031413058833177640960 
+ 120991848351738482367922176N
 \\+ 266735758564462825159262208N^2 + 
356790558744797070187560960N^3 + 322230995604339689974136832N^4\\ + 
206410863077042429645291520N^5 + 95396352174319203631759360N^6 + 
31338576528595789665009664N^7\\ + 6741729364322236678275072N^8 + 
 606442638553174662709248N^9 - 158967889827748034248704N^{10}\\ - 
 78160816104265974611968N^{11} - 16419585036974248984576N^{12} - 
     1908913019215816540160N^{13} \\- 62769211853172834304N^{14} + 
     19736004882625224704N^{15} + 4130688305419677696N^{16}\\ + 
     421867767284303872N^{17} + 25241612021992960N^{18} + 
693193236915968N^{19} \\- 18814089206912N^{20} 
- 2684704080320N^{21} - 
120121949760N^{22} - 2872757568N^{23}\\ 
- 35919232N^{24} - 178096N^{25}$

$Q_{20}(N)=    N^{20} (2 + N)^9 (4 + N)^5 (6 + N)^3 (8 + N)^3 (10 + N)
 (12 + N) (14 + N) (16 + N) (18 + N)$

\FL
$P_{21}(N)= 1351534773860942603511398400 + 
     6365460757030282723495772160N\\ 
+ 13733120620155454487896522752N^2 + 
     17929816694573858749176348672N^3\\ 
+ 15741441712440400461107822592N^4 + 
     9736866800557416285986619392N^5 \\
+ 4291917210346656516848746496N^6 + 
     1307842644179764469724872704N^7 \\
+ 238056916142751738992001024N^8 + 
     3543432139088928241090560N^9 \\
- 12741108714260109576372224N^{10} - 
     4295859766813135292465152N^{11}\\ 
- 755125851031606609051648N^{12} - 
     65023497347980126945280N^{13}\\ 
+ 2821567036764628910080N^{14} + 
     1727752072630205923328N^{15} \\
+ 264572837767576051712N^{16} + 
     22840865368902557696N^{17} \\
+ 1035703381014509568N^{18} - 
     6131507476388352N^{19} \\- 4493143518510080N^{20} 
- 338088589058432N^{21} - 
     14045533700352N^{22} \\
- 355361402880N^{23} - 5200818400N^{24} - 
     35916176N^{25} - 47360N^{26}$

$Q_{21}(N)=   N^{21}(2 + N)^9(4 + N)^5(6 + N)^3
(8 + N)^3(10 + N)(12 + N)
(14 + N)(16 + N)(18 + N)(20 + N)$

In particular for $N=0$ we have (in terms of the variable 
$\tilde \beta=\beta/N$):

$\chi(0,\tilde\beta) =1+ 4\tilde\beta
+ 12\tilde\beta^{ 2}+ 36\tilde\beta^{ 3}+ 
100\tilde\beta^{ 4}+ 284\tilde\beta^{ 5}+ 780\tilde\beta^{ 6}+ 
2172\tilde\beta^{ 7}+ 5916\tilde\beta^{ 8}+ 16268\tilde\beta^{ 9}+ 
44100\tilde\beta^{ 10}
+\\ 120292\tilde\beta^{ 11}+ 324932\tilde\beta^{ 12}+ 
881500\tilde\beta^{13 }
+ 2374444\tilde\beta^{ 14}+ 6416596\tilde\beta^{15 }+
17245332\tilde\beta^{16}+ 46466676\tilde\beta^{17}+
124658732\tilde\beta^{18 }+\\
335116620\tilde\beta^{19}+
897697164\tilde\beta^{20}+
2408806028\tilde\beta^{21}$.

 For $N=2$ we have:

$\chi(2,\beta) =   
1+ 2\beta+ 3\beta^{2 }+ 17/4\beta^{ 3}+
 11/2\beta^{ 4}+\\
 329/48\beta^{ 5}+ 529/64\beta^{ 6}+ 14933/1536\beta^{ 7}
+ 5737/512\beta^{ 8}+\\ 389393/30720\beta^{ 9}+ 
  2608499/184320\beta^{ 10}+ 3834323/245760\beta^{ 11}
+ 1254799/73728\beta^{ 12}\\
+ 84375807/4587520\beta^{ 13}+ 
  6511729891/330301440\beta^{ 14}
+ 66498259799/3170893824\beta^{ 15}+ \\
1054178743699/47563407360\beta^{ 16}+ 
  39863505993331/1712282664960\beta^{ 17}+
 19830277603399/815372697600\beta^{ 18}+ \\
  8656980509809027/342456532992000\beta^{19}+
  2985467351081077/114152177664000\beta^{20}+\\
  811927408684296587/30136174903296000\beta^{21}$

For $N=3$ we have:

$\chi(3,\beta) =  
1+ 4/3\beta+ 4/3\beta^{ 2}+ 56/45\beta^{ 3}+ 428/405\beta^{ 4}+\\
 1448/1701\beta^{ 5}+ 28048/42525\beta^{ 6}
+ 314288/637875\beta^{ 7}+ \\
  685196/1913625\beta^{ 8}+ 6845144/27064125\beta^{ 9}+
 1159405664/6630710625\beta^{ 10}+ \\
  643017322016/5430552001875\beta^{ 11}+
 915294455744/11636897146875\beta^{ 12}+\\
  12550612712128/244374840084375\beta^{ 13}+
 120892276630256/3665622601265625\beta^{ 14}+ \\
  3896992088570128/186946752664546875\beta^{ 15}+
 16982959056655084/1308627268651828125\beta^{ 16}+ \\
  23336075876557256/2930091359102578125\beta^{ 17}+ \\
  6224368647227625667744/1292302143475396569140625\beta^{ 18}+ \\
  22688150130310720609472/7897401987905201255859375\beta^{ 19}+\\  
120072005214715268585744/71076617891146811302734375\beta^{ 20}+\\
43219200109596671558015312/44138579710402169818998046875\beta^{21}$.

\section{ The second correlation moment}

The HT coefficients of the second correlation moment
$ \mu_{2}(N,\beta)=  \sum_{r=1}^\infty s_r(N) \beta^r $
 are:

$s_1(N)=4/N$

$s_2(N)=32/N^2  $

$s_3(N)= (328 + 160 N)/{(N^3 (2 + N))}$

$s_4(N)= (1408 + 640 N)/{(N^4 (2 + N))}$

$s_5(N)= (21728 + 14232 N + 2208 N^2)/{(N^5 (2 + N)(4 + N))}$

$s_6(N)= (156928 + 171072 N + 59840 N^2 + 6784 N^3)
/(N^6 (2 + N)^2 (4 + N))$

For the higher order coefficients $s_r(N)$ it is convenient to set 
$s_r(N)=P_r(N)/ {Q_r(N)}$  and to tabulate separately the numerator
$P_r(N)$ and the denominator $Q_r(N)$.

$P_{7}(N)=6487296 + 10904512 N + 7052384 N^2 + 2192384 N^3 
+ 328688 N^4 + 18944 N^5$

$Q_{7}(N)= N^8 (2 + N)^3 (4 + N) (6 + N)  $

$P_{8}(N)=21596160 + 34468352 N 
+ 21040128 N^2 + 6162816 N^3 + 878208 N^4 + 
     48640 N^5 $

$Q_{8}(N)=N^9 (2 + N)^3 (4 + N) (6 + N)$

$P_{9}(N)=560007168 + 912207616 N + 585628864 N^2 + 190951904 N^3 
+ 33905168 N^4 +  3117448 N^5 + 115616 N^6 $

$Q_{9}(N)= N^9 (2 + N)^3 (4 + N) (6 + N) (8 + N)$ 

$P_{10}(N)=14220853248 + 32437182464N + 31140450304 N^2 + 
16449182208 N^3 +  \\
     5251439360N^4 + 1045045888 N^5 + 127390272 N^6 + 8712896 N^7 + 
     255616 N^8$

$Q_{10}(N)=N^{10}(2 + N)^4(4 + N)^2(6 + N)(8 + N)$

$P_{11}(N)=3549643407360 + 10748529770496 N + 14330317561856 N^2 
+  \\ 11099228633088 N^3 + 5553033387520 N^4 
+ 1888211571200 N^5 +  \\
     446700183296 N^6 + 73788019072 N^7 + 8364424672 N^8 
+ 620464608 N^9 +  \\
     27094944 N^{10} + 526848 N^{11}$

$Q_{11}(N)=   N^{11} (2 + N)^5 (4 + N)^3 (6 + N)(8 + N)(10 + N)$

$P_{12}(N)=10920048721920 + 31897824264192 N  
+ \\ 40878320844800 N^2 + 
     30320450846720 N^3 + 14475616055296 N^4 
+ 4684012402688 N^5 +  \\
     1053225265152 N^6 + 165529253888 N^7 + 17912071424 N^8 + 
     1274573696 N^9 + \\ 53696896 N^{10} + 1013248 N^{11}$

$Q_{12}(N)= N^{12}(2 + N)^5(4 + N)^3(6 + N)(8 + N)(10 + N)$ 

$P_{13}(N)=398311403028480 + 1152646176964608N 
+ 1467863723606016N^2 +  \\
  1086884435984384N^3 + 521655212892160N^4 + 171508785347072N^5 + 
  39822249767936N^6 + \\ 6624256364416N^7 + 788536224640N^8 + 
  65778403648N^9 + 3655303744N^{10} 
+ 121407936N^{11} + \\ 1818880N^{12}$

$Q_{13}(N)=N^{13}(2 + N)^5(4 + N)^3(6 + N)(8 + N)(10 + N)(12 + N)$

$P_{14}(N)=14381419069440000 + 49571261401006080N 
+ 76559245123780608N^2 +  \\
     70130377503539200N^3 + 42548875902058496N^4 
+ 18101157266890752N^5 + \\ 
     5581472628355072N^6 + 1272436773582848N^7 
+ 216726437780480N^8 + 
     27616170103808N^9 \\ 
+ 2605954492416N^{10} + 177259542528N^{11} + 
     8230177408N^{12} + 233423296N^{13} + 3045888N^{14}$

$Q_{14}(N)=N^{14}(2 + N)^6(4 + N)^3(6 + N)^2(8 + N)(10 + N)(12 + N)$

$P_{15}(N)=7205314087322910720 + 29316772904369651712N  \\
	+ 54344822673767399424N^2 + 
     60875749209938067456N^3 + 46110757538166996992N^4 +  \\
     25056228677107990528N^5 + 10119823516255248384N^6 + 
     3107917444672675840N^7  \\+ 736807917027831808N^8 + 
     136144584030477312N^9 + 19700616020237824N^{10} +  \\
     2231029399055360N^{11} 
+ 196196174024448N^{12} + 13160346699904N^{13} + 
     651384063232N^{14} + \\ 22420625856N^{15} 
	+ 478573168N^{16} + 4759552N^{17}$

$Q_{15}(N)=N^{15}(2 + N)^7(4 + N)^3
(6 + N)^3(8 + N)(10 + N)(12 + N)(14 + N)$

$P_{16}(N)=14215559923441336320 + 57338919520296763392N +  \\
     105267543605621293056N^2 + 116619791028818280448N^3 + 
     87189583554941026304N^4  \\+ 46640416019329581056N^5 + 
     18481743700474920960N^6 + 5546746731056758784N^7 +  \\
     1279666393060327424N^8 + 229248311936622592N^9 + 
     32092209212850176N^{10} + \\ 3517151947112448N^{11} 
+ 300336273417216N^{12} + 
     19680328220160N^{13}  
+  958823550720N^{14} + \\ 32735074432N^{15} + 
     697558656N^{16} + 6957056N^{17}$

$Q_{16}(N)=N^{16}(2 + N)^7(4 + N)^4
(6 + N)^2(8 + N)(10 + N)(12 + N)(14 + N)$

$P_{17}(N)=16044762062435301457920 + 70470061100481487306752N +  \\
     142655241108775004798976N^2 + 176761774102693835440128N^3 +  \\
     150198190696943073624064N^4 + 92960797041997201276928N^5 +  \\
     43465361502957674430464N^6 + 15723355668874622926848N^7 +  \\
     4473127620779360665600N^8 + 1012537718607308627968N^9 + 
     183938871531499802624N^{10} + \\ 26986540252155746304N^{11} + 
     3211089966275868672N^{12} + 310291779078352896N^{13} +  \\
     24273059658657280N^{14} + 1521460496772864N^{15} 
+ 74832996108608N^{16} +  \\
     2783265847904N^{17} + 73449080400N^{18} 
+ 1221110008N^{19} + 9573792N^{20}$

$Q_{17}(N)= N^{17}(2 + N)^7(4 + N)^5(6 + N)^3(8 + N)(10 + N)(12 + N)
(14 + N)(16 + N)$

$P_{18}(N)=750029539465981990010880 + 3671118091050821760319488N +  \\
 8335725431444873454551040N^2 + 11665373867250167725424640N^3 +  \\
 11276915336572927456837632N^4 + 8001145549336838213206016N^5 +  \\
 4322734533275908527095808N^6 + 1821577961516403890585600N^7 +  \\
 608650745059201940193280N^8 + 163152472998781338124288N^9 +  \\
 35389047160612482121728N^{10} + 6253363024804357767168N^{11} +  \\
 905249991372291489792N^{12} + 107885738708892270592N^{13} +  \\
 10627036460386877440N^{14} + 866704728760623104N^{15} +  \\
 58356430103347200N^{16} + 3208488174566912N^{17} 
+ 140753928694016N^{18} + \\ 
  4731693506944N^{19} + 113947558464N^{20} + 
1740125248N^{21} +\\ 12590720N^{22}$

$Q_{18}(N)=N^{18}(2 + N)^8(4 + N)^5(6 + N)^3(8 + N)^2(10 + N)
(12 + N)(14 + N)(16 + N)$

$P_{19}(N)=627690146360050712617943040 
+ 3421601811563754334552326144N + \\ 
     8712245471445327025300045824N^2 
+ 13774632993059184247962599424N^3 +  \\
     15165775997343900382418436096N^4 
+ 12361884192755007745538654208N^5 +  \\
     7744311739948445340207677440N^6 
+ 3821697127832819813536759808N^7 +  \\
     1511074094374096761920684032N^8 
+ 484537143774775297319108608N^9 +  \\
     127124521376337134571356160N^{10} 
+ 27474245822576876554944512N^{11} + \\ 
     4918326246120771429597184N^{12} 
+ 733003199414399719473152N^{13} +  \\
     91428299627176181227520N^{14} 
+ 9600359324677991481344N^{15} + \\ 
     853907491123648274432N^{16} 
+ 64659218756176879616N^{17} + 
     4171870520224442880N^{18} + \\ 227737937559751168N^{19} + 
     10327555689201152N^{20} + 377014913696896N^{21} + \\
 10570062577056N^{22} + 
     212003127744N^{23} +  2690873184N^{24} + 16165376N^{25}$

$Q_{19}(N)= N^{19}(2 + N)^9(4 + N)^5(6 + N)^3(8 + N)^3(10 + N)
(12 + N)(14 + N)(16 + N)(18 + N)$

$P_{20}(N)=1815178094017325774709719040 + 
   9666760450785608126844370944 N\\ 
+ 24005039860199234508542705664 N^2 + 
   36939925977092556006616989696 N^3\\ 
+ 39491506382084850077046669312 N^4 + 
   31171265041933192528668917760 N^5 
\\+ 18848811341049400534841163776 N^6 + 
   8944169357858645440582385664 N^7 
\\+ 3385359998666615144499380224 N^8 + 
   1033678218058491052840452096 N^9 
\\+ 256651703697572047547793408 N^{10} + 
   52123814454053216103956480 N^{11}\\ 
+ 8701745257544758809853952 N^{12} + 
   1200498356471982676639744 N^{13} 
\\+ 137873846191171430711296 N^{14} + 
   13328241986441571500032 N^{15} 
\\+ 1101370801394711199744 N^{16} + 
   79107808271490596864 N^{17} 
\\+ 4985190283611369472 N^{18} + 
   273323887975057408 N^{19}\\ 
+ 12659942377698304 N^{20} + 
   473471116241408 N^{21} 
+ 13509163363584 N^{22}\\ + 272911256960 N^{23} + 
   3455684224 N^{24} + 20575744 N^{25}$

$Q_{20}(N)= N^{20} (2 + N)^9 (4 + N)^5 (6 + N)^3 (8 + N)^3 (10 + N)
	 (12 + N) (14 + N) (16 + N) (18 + N)$

$P_{21}(N)=104523299485865383076403609600 
+ 548601456372456927606016573440N \\
+ 1341859981319362312024199528448N^2 
+ 2032492529321839239165797990400N^3\\ + 
 2137001952081648448593343807488N^4 
+ 1657243331724148071231538593792N^5\\ + 
 983339698515439540272495591424N^6 
+  457150573406528768021251489792*^7\\ + 
 169173985424612010389621702656N^8 
+  50367854640918058439851114496N^9\\ + 
 12150736632633303934995791872N^{10} 
+  2386446463364998372987371520N^{11}\\ 
+ 383008009888335829595848704N^{12} 
+      50455608541331884025479168N^{13}\\ 
+ 5501372995701701085822976N^{14} 
+      505026346896678580174848N^{15}\\ 
+ 40256133352226331713536N^{16} 
+      2903228052786320875520N^{17} 
+ 194825662990468552704N^{18}\\ 
+      12002431296774673920N^{19} 
+ 645176662272934912N^{20} 
+      28519050415695488N^{21} \\
+ 983723470664448N^{22} 
+ 25174415011840N^{23} +      447917101984N^{24}
+ 4947929456N^{25}\\ + 25626368N^{26}$

$Q_{21}(N)=N^{21}(2 + N)^9(4 + N)^5(6 + N)^3(8 + N)^3(10 + N)
(12 + N)(14 + N) (16 + N)(18 + N)(20 + N)$

In particular for $N=0$ we have (in terms of the variable 
$\tilde \beta=\beta/N$):

$\mu_{2}(0,\tilde\beta) =  4\tilde \beta+ 32\tilde \beta^{ 2}+
 164\tilde \beta^{ 3}+ \\
704\tilde \beta^{ 4}+ 2716\tilde \beta^{ 5}
+ 9808\tilde \beta^{ 6}+\\
 33788\tilde \beta^{ 7}
+ 112480\tilde \beta^{ 8}+ 364588\tilde \beta^{ 9}+\\
 1157296\tilde \beta^{ 10}+ 3610884\tilde \beta^{ 11}+ 
  11108448\tilde \beta^{ 12}+\\ 33765276\tilde \beta^{ 13}+
 101594000\tilde \beta^{ 14}+ 302977204\tilde \beta^{ 15}+ \\
896627936\tilde \beta^{ 16}+ 2635423124\tilde \beta^{ 17}+ 
  7699729296\tilde \beta^{ 18}+ \\ 22374323436\tilde \beta^{ 19}+
64702914336\tilde\beta^{ 20}+
186289216332\tilde\beta^{ 21}$.

For $N=2$ we have:

$\mu_{2}(2,\beta) = 
 2\beta+ 8\beta^{2}+ 81/4\beta^{3}+ 42\beta^{4}+ 3689/48\beta^{5}+ 
6193/48\beta^{6}+\\ 312149/1536\beta^{7}+ 19499/64\beta^{8}+
 13484753/30720\beta^{9}+ 
  28201211/46080\beta^{10}+\\ 611969977/737280\beta^{ 11}+ 
101320493/92160\beta^{ 12}+ 58900571047/41287680\beta^{ 13}+\\
  3336209179/1835008\beta^{ 14}
+ 1721567587879/754974720\beta^{ 15}+\\
 16763079262169/5945425920\beta^{ 16}+ 
  5893118865913171/1712282664960\beta^{ 17}+\\
 17775777329026559/4280706662400\beta^{ 18}+
  1697692411053976387/342456532992000\beta^{ 19}+\\
41816028466101527/7134511104000\beta^{ 20}+
206973837048951639371/30136174903296000\beta^{21}$.

For $N=3$ we have:

$\mu_{2}(3,\beta) = 
 4/3\beta+  32/9\beta^{ 2}+ 808/135\beta^{ 3}+ 3328/405\beta^{ 4}+\\
 84296/8505\beta^{ 5}+ 1391872/127575\beta^{ 6}+ 
  21454864/1913625\beta^{ 7}+\\ 62634752/5740875\beta^{ 8}+
 1923459304/189448875\beta^{ 9}+ \\
  25854552704/2841733125\beta^{ 10}+
 23813358832544/3016973334375\beta^{ 11}+ \\
  180728998866176/27152760009375\beta^{ 12}+\\ 
148615553292224/27152760009375\beta^{ 13}+ \\
  16130002755113536/3665622601265625\beta^{ 14}+\\
 647957301000434704/186946752664546875\beta^{ 15}+ \\
 420771056234707712/157035272238219375\beta^{ 16}+ \\
 15946537289290290889672/7832134202881191328125\beta^{ 17}+ \\
 655875829145233998723328/430767381158465523046875\beta^{ 18}+ \\
 26576776651881149990183488/23692205963715603767578125\beta^{ 19}+\\
24824955666477074672626688/30461407667634347701171875\beta^{ 20}+\\
318405686546338787648327888/544920737165458886654296875\beta^{21}$.

\section{ The second field derivative of the susceptibility}

The HT coefficients of the second field derivative of the 
susceptibility
$ \chi_{4}(N,\beta )=  
{\frac {1} {N}} \Big( -2 
+ \sum_{r=1}^\infty d_r(N) \beta^r \Big)$  are:

$d_1(N)=-32/N$

$d_2(N)=-8(64 + 35N)/N^2(2 + N)$

$d_3(N)=-256(12 + 7N)/N^3(2 + N)$

$d_4(N)=-8(15328 + 20856N + 8952N^2 + 1169N^3)/N^4(2 + N)^2(4 + N)$ 
 
For the higher order  coefficients $d_r(N)$ it is convenient to set 
$d_r(N)=P_r(N)/ {Q_r(N)}$  and to tabulate separately the numerator
$P_r(N)$ and the denominator $Q_r(N)$.

$P_5(N)= -64(8568 + 11666N + 5033N^2 + 656N^3)$

$Q_5(N)=N^5(2 + N)^2(4 + N)$
 
$P_6(N)=-32(848448 + 1708560N + 1320504N^2 + 483386N^3 + 82492N^4 + 
       5229N^5)$

$Q_6(N)=N^6(2 + N)^3(4 + N)(6 + N)$

$P_7(N)= -256(413760 + 819248N + 623078N^2 + 224569N^3 + 37724N^4 + 
       2360N^5)$

$Q_7(N)=N^7(2 + N)^3(4 + N)(6 + N)$

$P_8(N)=-8(3160154112 + 8880870400N 
+ 10579850240N^2 + 6952508224N^3 + 
       2745320192N^4 +\\ 664622096N^5 + 96137184N^6 + 7591150N^7 + 
       250437N^8)$

$Q_8(N)=N^8(2 + N)^4(4 + N)^2(6 + N)(8 + N)$

$P_9(N)=-64(1423288320 + 3917961472N 
+ 4569267968N^2 + 2938300752N^3 + \\
       1135415360N^4 + 269158808N^5 + 38175104N^6 + 2962181N^7 + 
       96280N^8$

$Q_9(N)=N^9(2 + N)^4(4 + N)^2(6 + N)(8 + N)$

$P_{10}(N)=-32(797183508480 + 2819876773888N + 4414993874944N^2 + \\
       4031840569344N^3 + 2382096324608N^4 + 954513246240N^5 + \\
 264332662544N^6 + 50539833080N^7 + 6533960024N^8 +  543728966N^9 \\
+ 26196458N^{10} + 553189N^{11})$

$Q_{10}(N)=N^{10}(2 + N)^5(4 + N)^3(6 + N)(8 + N)(10 + N)$

$P_{11}(N)=-256(340762460160 + 1180158980096N + 1807166870528N^2 + \\
    1612510528512N^3 + 930184344000N^4 + 363788558192N^5 + \\
 98342090128N^6 + 18368180928N^7 + 2323043092N^8 + 189484599N^9 + \\
       8969234N^{10} + 186536N^{11})$

$Q_{11}(N)=N^{11}(2 + N)^5(4 + N)^3(6 + N)(8 + N)(10 + N)$

$P_{12}(N)=-32(1315703150346240 + 5438217813295104N +\\
 10169455328329728N^2 + 
  11390757326471168N^3 + 8529038337966080N^4 + \\
 4512300585885440N^5 + 1738422710398464N^6 
+ 495320225789376N^7 + \\
 104883376470528N^8 + 16428534330480N^9 + 1874379299088N^{10} + \\
 151083753396N^{11} + 8133230252N^{12} 
+ 261683156N^{13} + 3795185N^{14})$

$Q_{12}(N)=N^{12}(2 + N)^6(4 + N)^3(6 + N)^2(8 + N)(10 + N)(12 + N)$

$P_{13}(N)=-512(270976473169920 + 1096725428969472N +\\
 2005753217814528N^2 + 
 2194539397009408N^3 + 1603311500740608N^4 +\\ 826889848699616N^5 + 
 310370441728160N^6 + 86142909672256N^7 +\\ 17775181813000N^8 + 
 2715815783098N^9 + 302703436000N^{10} +\\ 23883276726N^{11} + 
 1261330187N^{12} + 39906694N^{13} + 570432N^{14})$

$Q_{13}(N)=N^{13}(2 + N)^6(4 + N)^3(6 + N)^2(8 + N)(10 + N)(12 + N)$

$P_{14}(N)=-64(4729747131240284160 + 23387807447028596736N + \\
       53391491102489444352N^2 + 74720767414227566592N^3 + \\
       71798269494072377344N^4 + 50281426991956893696N^5 + \\
       26590502100001992704N^6 + 10856939418704423424N^7 + \\
       3470103039376496128N^8 + 874999045177634816N^9 + \\
       174555747726402112N^{10} + 27499523899040704N^{11} + \\
       3397617398241328N^{12} + 324891160500608N^{13} \\
	+ 23521314618812N^{14} + 
   1244052846554N^{15} + 45255618107N^{16} +\\ 1009996430N^{17} + 
       10401518N^{18})$

$Q_{14}(N)=N^{14}(2 + N)^7(4 + N)^4(6 + N)^3(8 + N)
(10 + N)(12 + N)(14 + N)$

$P_{15}(N)=-512(1895033761431552000 + 9190498448855531520N + \\
       20555882862533148672N^2 + 28154602101819899904N^3 + \\
       26448192225888174080N^4 + 18088913492373442560N^5 + \\
       9333532357592239104N^6 + 3715399595069458688N^7 + \\
       1157147824100880000N^8 + 284257647952320768N^9 + \\
       55255671532695936N^{10} + 8487432924662416N^{11} + \\
  1023523926260184N^{12} + 95669381062712N^{13} 
+ 6782408485072N^{14} + \\
 351973151712N^{15} + 12588857511N^{16} 
+ 276796330N^{17} + 2813856N^{18})$

$Q_{15}(N)=N^{15}(2 + N)^7(4 + N)^4(6 + N)^3(8 + N)(10 + N)(12 + N)*
     (14 + N)$

$P_{16}(N)=-8(393553748694956601507840 + \\
  2238797016999938557476864N + 5952616821086089377742848N^2 + \\
  9835930646135145322512384N^3 + 11329546912664804682366976N^4 + \\
  9673321212788105516417024N^5 + 6356189412895881926017024N^6 + \\
  3294382649182089746055168N^7 + 1369622314378844679962624N^8 + \\
 462055197809790960812032N^9 + 127470718364764484878336N^{10} + \\
 28892571739468927709184N^{11} + 5391343629949283622912N^{12} + \\
 827734092308104586240N^{13} + 104208728930679520256N^{14} + \\
 10686907762867520768N^{15} + 883185743012455424N^{16} + \\
 57860284710989984N^{17} + 2931823000694800N^{18} + \\
 110628321153680N^{19} + 2921354278696N^{20} + 48084058522N^{21} + \\
 370591789N^{22})$

$Q_{16}(N)=N^{16}(2 + N)^8(4 + N)^5(6 + N)^3(8 + N)^2(10 + N)
(12 + N)(14 + N)(16 + N)$

$P_{17}(N)=-64(154328333132805848432640 + \\
  862171270522562364309504N + 2249151812525864836399104N^2 + \\
  3642762103032724416626688N^3 + 4108543937641919350308864N^4 + \\
  3431288500060965921882112N^5 + 2203091748524324305240064N^6 + \\
  1114625140117096539160576N^7 + 451935655809348535320576N^8 + \\
  148577932801385167507456N^9 + 39921739674229256839168N^{10} + \\
  8810453666427880972288N^{11} + 1600832358513749923840N^{12} + \\
  239426089299260737792N^{13} + 29389521370216614400N^{14} + \\
  2942373642638070784N^{15} + 237767376253506048N^{16} + \\
15259439999351088N^{17} + 758951651849152N^{18} 
+ 28166343349968N^{19} + \\
       732948825144N^{20} + 11909447319N^{21} + 90757800N^{22})$

$Q_{17}(N)=N^{17}(2 + N)^8(4 + N)^5(6 + N)^3(8 + N)^2(10 + N)
(12 + N)(14 + N)(16 + N)$

 If we compute $N\chi_{4}(N,\tilde\beta)$  for $N=0$ we get:

$-2
-32\tilde\beta -256\tilde\beta^{ 2} -1536\tilde\beta^{ 3}
 -7664\tilde\beta^{ 4} 
-34272\tilde\beta^{ 5}\\ -141408\tilde\beta^{ 6}
 -551680\tilde\beta^{ 7} -2057392\tilde\beta^{ 8}
 -7412960\tilde\beta^{ 9} 
  -25949984\tilde\beta^{ 10}\\ -88740224\tilde\beta^{ 11} 
-297422768\tilde\beta^{ 12} -980094304\tilde\beta^{ 13}\\
 -3182108192\tilde\beta^{ 14} -10199619200\tilde\beta^{ 15} \\
  -32321471824\tilde\beta^{ 16} -101396444832\tilde\beta^{ 17}$.

For $N=1$ we have:

$\chi_{4}(1,\beta) =-2 -32\beta 
-264\beta^{2 }-4864/3\beta^{3 } -8232\beta^{4} 
-553024/15\beta^{ 5}\\ 
-2259616/15\beta^{ 6} -180969728/315\beta^{ 7} 
 -217858792/105\beta^{ 8} -20330135104/2835\beta^{ 9}\\
 -5377792736/225\beta^{ 10} -12048694416128/155925\beta^{ 11} 
 -3450381618464/14175\beta^{ 12}\\ 
-4559524221383168/6081075\beta^{ 13} 
 -32137492094329792/14189175\beta^{ 14}\\
 -4294238083842489856/638512875\beta^{ 15} 
 -66447301472480024/3378375\beta^{ 16} \\
-615947855084824982464/10854718875\beta^{ 17}$

 For $N=2$ we have:

$\chi_{4}(2,\beta) = -1
-8\beta -67/2\beta^{2 } -104\beta^{3 } -12775/48\beta^{ 4}
 -1790/3\beta^{ 5}\\ -931367/768\beta^{ 6} -109691/48\beta^{ 7} 
  -93380347/23040\beta^{ 8} -157557481/23040\beta^{ 9}\\
 -8158367639/737280\beta^{ 10} -1061565359/61440\beta^{ 11} \\
 -1296061504531/49545216\beta^{ 12} 
-477508721605/12386304\beta^{ 13} \\
  -439777014509471/7927234560\beta^{ 14} 
-245861555567/3145728\beta^{ 15} \\
  -462463818305826161/4280706662400\beta^{ 16} 
-314178739246240667/2140353331200\beta^{ 17}$.

For $N=3$ we have:

$\chi_{4}(3,\beta) = 
-2/3 -32/9\beta -1352/135\beta^{2 } -2816/135\beta^{3 }
 -1520216/42525\beta^{ 4}\\  -12992/243\beta^{ 5} 
  -3070624/42525\beta^{ 6} -516883712/5740875\beta^{ 7}\\
 -697726412216/6630710625\beta^{ 8} 
  -331122359872/2841733125\beta^{ 9}\\
 -478066539947936/3878965715625\beta^{ 10} 
  -185574375218432/1481059636875\beta^{ 11}\\
 -150342773008769632/1221874200421875\beta^{ 12} 
  -429508071453349376/3665622601265625\beta^{ 13} \\
  -710293648879287815872/6543136343259140625\beta^{ 14} \\
  -1925804659821618529792/19629409029777421875\beta^{ 15} \\
  -2872493310184686424756616/33135952396805040234375\beta^{ 16} \\
  -32321239221821813512332352/430767381158465523046875\beta^{ 17}$.

\begin{figure}
\figure {Fig. 1. Numerical estimates of 
the critical exponent $\gamma(N)$ of the 
susceptibility, of the exponent $\nu(N)$ of the correlation 
length and  of the exponent $2\Delta(N)$ from the 
second field derivative of the susceptibility, as computed for
 $ -2 \leq  N < 2$ by the method described in Section 3. Our  
results are represented by (the centers of)  
circles for the exponent $\nu(N)$, 
 squares for the exponent 
$\gamma(N)$ and  triangles for twice the gap exponent $\Delta(N)$ 
and they are compared
 with the corresponding exact 
formulas   (\ref{niennu}) (dashed line), 
(\ref{niengam}) (continuous line) 
and  (\ref{ipers}) (dot-dashed line) 
respectively. Whenever no error bars appear, they are smaller than
the data point.} \label{fig1}
\end{figure}
\begin{figure}
\figure{Fig. 2. Our estimates of the dimensionless renormalized 
coupling constant $\hat g_r(N)$ for various 
values of $0 \leq N \leq \infty $
 compared to some results from a recent 
MonteCarlo(MC) cluster computation\cite{kim} for $N=2$ and $3$, 
 to a field theoretic estimate in the case $N=3$\cite{falcioni} 
and  to other  HT estimates
\cite{nickelsharpe,campo}. 
For comparison, we have also plotted the large $N$ 
asymptotic behavior(\ref{grn}).}\label{fig2}

\figure{Fig. 3. The quantity $A_r(N) \equiv (N/2)^r  a_r(N)$  
 versus the order $r$ for various fixed values of $N$. 
This normalization of the expansion coefficients $a_r(N)$
 of the susceptibility has been chosen in order to make the 
plots for different values of $N$  more easily 
comparable.  We have also interpolated 
the data points  by smooth continuous curves only 
to guide the eye.}\label{fig3}
\end{figure}

\begin{figure}
\figure{Fig. 4. The real part 
(circles) and the imaginary part (triangles)
of the nearest singularity $\tilde \beta_n$ 
of the susceptibilty $\chi(N,\beta)$ 
in the complex $\tilde\beta= \beta/N$ plane  plotted 
as  functions of $x\equiv 1-1/N$   in the range $1 \leq N < \infty$.
For $ x \protect \lesssim 0. 52$, the singularity 
$\tilde \beta_n$ is still a real
critical point, but for larger $x$ 
it splits into a pair of complex conjugate 
singularities which move into the complex plane and,
 as $N \rightarrow \infty$,  tend to the limiting
points $\tilde \beta_{\pm} \simeq 0.32162(1 \pm i)$.
Here we have plotted only the trajectory of the 
singularity in the first quadrant of  
 the complex $\tilde\beta$ plane. }\label{fig4}
\end{figure}
\begin{figure}
\figure{Fig. 5. The quantity  $B_{\chi}(N,\tilde \beta)$ 
 defined by eq.(\ref{bizeroxi}) versus $\tilde \beta$  
for  $N=7$ (lower set of curves) and $N=4$ (upper set)
 showing the existence of a stationary point 
$\tilde \beta_s$ at which we estimate $b_0(N)$. We have plotted
 PA's of $B_{\chi}(N,\tilde \beta)$  which use at least 15 HT series 
coefficients and with a  difference between the 
degrees  of numerator and 
denominator  not larger than 4. }\label{fig5}
\end{figure}

\begin{figure}
\figure{Fig. 6.  Numerical estimates 
of $b_0(N)$ obtained starting from 
the quantities 
$B_{\chi}(N,\tilde \beta_s)$ (triangles), 
$B_{\xi}(N,\tilde \beta_s)$ 
(circles) and 
$B_{4}(N,\tilde \beta_s)$ (squares) are plotted
versus $N$. Only for graphical 
convenience the estimates have been 
computed for three different sets of noninteger values of $N$.
   They are  compared  to 
the expected value  (eq. \ref{betagamma}) 
represented by the continuous line.}\label{fig6}
\end{figure}
\begin{figure}
\figure{Fig. 7. Our numerical 
estimates of $c_{\xi}(N)$ ( triangles), 
$c_{\chi}(N)$ (circles) and $c_r(N)$ (squares)  
are compared to the  exact (\ref{bethe}) or conjectured
\cite{caralat95}  formulas (continuous lines), 
 and  to their $1/N$ 
expansions (dashed  lines).
 Please notice that for graphical 
convenience we have shifted 
upwards by 0.1 the data for $c_{\chi}(N)$
 and have scaled down by a factor .25 the data for $c_r(N)$. 
The actual numerical values  of these quantities are 
listed in Table 1.}\label{fig7}
\end{figure}

\widetext
\begin{table}
\caption{ Our central estimates 
of the universal constants $c_{\xi}(N)$, 
$c_{\chi}(N)$, $c_r(N)$   and of $g_r(N)$ 
for various values of  $N$. We have indicated only the 'statistical' 
uncertainty.
 Beside our estimates of $c_{\xi}(N)$, 
$c_{\chi}(N)$, $c_r(N)$,  we have reported 
in square parentheses the predicted value 
from eq. (\ref{bethe}) or from
the conjectured formula of Ref. \protect\cite{caralat95}. }
\begin{tabular}{ccccc}
$N$ &  $c_{\xi}(N)$ &$c_{\chi}(N)$ & $c_r(N)$& $ g_r(N)$  \\
\hline
0&  &    &                                        &10.53(2) \\
\hline
1&  &    &                                       &14.693(4) \\
\hline
2&  &    &                                          &18.3(2) \\
\hline
3.5& 0.028(8) [0.0273] &0.021(2)[0.0266] & 2.379(9)[2.45] &20.4(1) \\
\hline
4 &0.039(1) [0.0416]  &0.034(1)[0.0385] & 1.964(9)[2.007]&20.9(1) \\
\hline
5 &0.065(1) [0.0652] &0.059(3)[0.0600] &1.606(9)[1.624]&21.6(2) \\
\hline
6 &0.084(2) [0.0826] &0.077(4)[0.0776] &1.443(8)[1.452]& 21.9(3)\\
\hline
8 &0.106(2) [0.1054]  &0.1035(8)[0.1027] &1.290(9)[[1.291]&22.5(4)\\
\hline
10 &0.121(2) [0.1195] &0.1212(5)[0.1194] &1.213(5)[1.215]& 22.8(6)\\
\hline
12 &0.130(2) [0.1290] &0.134(2)[0.1312]  &1.169(5)[1.171] &23.1(6)\\
\hline
14 &0.137(2) [0.1358] &0.143(1)[0.1399]  &1.141(3)[1.141] &23.3(6)\\
\end{tabular}
\end{table}


\begin{references}
\bibitem{bcxy21} Series $O(\beta^{14})$ for the 
triangular lattice  have been computed and analysed in \cite{bcxy22}.
\bibitem{nota} The oscillations in sign 
of the HT expansion coefficients 
of $\chi$ in  the spherical model 
were pointed out long ago \cite{dw} (without further explanation)
 to occur both in the case
 of the square lattice and
of the triangular lattice.
 The coefficients first become negative 
at 18th  term in the former case
and at 13th in the latter.
This remark has been ignored in most later works. For example
Ref.\cite{parisi} is based on 
the conjecture that the susceptibility
 coefficients are all positive for any $N \geq 3$.
 Even in the recent literature, one  still 
finds  the  statement
 that the HT coefficients of the 
susceptibility for the $O(N)$ models 
are all positive "by inspection" 
\cite{reisz}, a correct statement being, of course, only 
that the susceptibility is a positive quantity.
\end{references}
\end{document}